%% file: smatrix.tex
\documentclass[aps,pre,twocolumn,showpacs]{revtex4}
\usepackage{graphics,color}
\usepackage{amsmath,amssymb}
\usepackage{wasysym,stmaryrd}

\newcommand{\dr}{\textrm{d}}
\newcommand{\ir}{\textrm{i}}
\newcommand{\er}{\textrm{e}}

\newcommand{\dint}{\int \!\!\!\! \int}

\begin{document}
\today
\title{Complete $S$-matrix in a microwave cavity at room temperature}
\author{J\'er\^ome Barth\'elemy, Olivier Legrand, Fabrice Mortessagne}
\affiliation{Laboratoire de Physique de la Mati\`ere Condens\'ee,
CNRS UMR 6622,\\  Universit\'e de Nice-Sophia Antipolis, 06108 Nice,
France}

\begin{abstract}
We experimentally study the widths of resonances in a two-dimensional
microwave cavity at room temperature. By developing a model for the
coupling antennas, we are able to discriminate their contribution from those
of ohmic losses to the broadening of resonances. Concerning ohmic losses, we
experimentally put to evidence two mechanisms: damping along propagation
and absorption at the contour, the latter being responsible for variations
of widths from mode to mode due to its dependence on the spatial
distribution of the field at the contour. A theory, based on an $S$-matrix formalism, is given for these variations. It  is successfully validated through measurements of several hundreds of resonances in a rectangular cavity.
\end{abstract}


\maketitle
 
\section{Introduction}
In the field of Quantum Chaos, microwave experiments have proved to yield
very important breakthroughs in providing versatile analog models of quantum
systems in the domain of classical electromagnetic waves\cite{Stoeckmannbook}. Room temperature experiments 
have opened the way\cite{Stoeckmann_1990} rapidly followed by experiments in
superconducting cavities\cite{Graef_1992}. In a first stage, studies have mainly
been concerned with the verification of predictions issued from Random
Matrix Theory or from semiclassical approaches regarding spectral
fluctuations. Losses, which were originally absent from theoretical models,
were seen as severe drawbacks in the seminal experiments, especially for an
accurate analysis of resonance frequencies (see e.g. \cite{Stoeckmann_1990, Graef_1992, Haake_1991}). The first account of resonance
widths observed in superconducting cavities was related to coupling losses
in the absence of ohmic losses and measuring widths essentially amounted to
measuring intensities at the locations of few antennas\cite{alt_1995}.
During the last decade, the great flexibility of microwave cavities has led to an important
diversification of geometries and configurations in order to investigate the spectral correlations and the spatial distribution of the field, in closed or open, disordered and/or
chaotic cavities (see \cite{Stoeckmannbook} for a review). Nevertheless, until recent years, the impact of the different loss mechanisms, present in these systems, on their spatial or spectral
statistical properties had obtained very little consideration. Indeed, as
long as losses are weak, resonances can be viewed as isolated. On the
contrary, for increasing damping, resonances are no longer easily
distinguished due to modal overlapping and the very description of the wave
system in terms of modes loses its pertinence. Since the seminal papers by Ericson in nuclear physics\cite{Ericson_1963} and by Schroeder in room acoustics\cite{Schroeder_1962b}, the regime of large modal overlap has been abundantly studied in the context of quantum chaos\cite{Verbaarschot_1985, Schafer_2003, Rozhkov_2003}. 

 The question of
intermediate modal overlap for which resonances can be distinguished but
broadening is no longer negligible is essentially open as yet (see the
excellent review \cite{Dittes_2000}). In the present paper we propose to help pave the way of a more complete understanding of microwave cavities at room temperature by accounting for the presence of essentially two kinds of loss mechanisms, namely ohmic damping at the boundaries and coupling to the outside through antennas. To be able to separate their respective contributions to the broadening of resonances, a thorough
analysis is required of the way the wavefunctions are spatially distributed throughout the cavity.

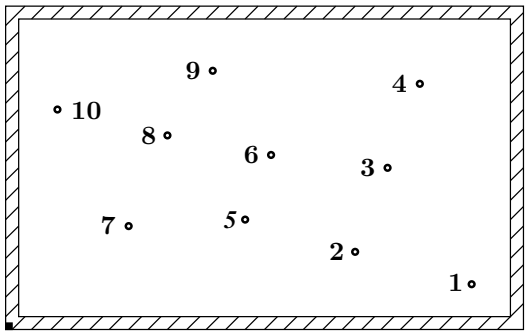
\begin{figure}
\begin{center}
\input{antenne_pos2.pstex_t}
\end{center}
\caption{Schematic view of the microwave cavity with locations of antennas.}
\label{cavity}
\end{figure}

The cavity we have actually used for our experiments is composed of two rectangular OFHC copper plates between which a copper rectangular frame is sandwiched. The rectangular frame has been machined as one piece and serves as the contour of the cavity. The cavity may thus be viewed as the slice of a rectangular waveguide closed at both \emph{ends}, with contour $\mathcal{C}$ of length $L=2.446\,$m, section $\mathcal{S}$ of area $A=0.3528\,$m$^2$ and thickness $d=5\,$mm. As long as the wavelength $\lambda$ is larger than $d$, the boundary conditions in the $z$ direction (perpendicular to the top and bottom plates)  only admit Transverse Magnetic (TM) two-dimensional (2D) modes. The whole structure is tightly screwed and 10 holes have been drilled through one of the plates to introduce 10 antennas, which protrude a length $l$ into the cavity. The antennas are monopolar with SMA connectors which are commonly used in the frequency range from 0 to 18\,GHz. The positions of the antennas are displayed on figure \ref{cavity}. For a measurement, only one antenna at a time is used as a microwave emitter and another (in transmission) or the same (in reflection) as a receiver. The other unused antennas are terminated by 50\,$\Omega$ loads so that all antennas behave the same way regarding the losses they imply. These antennas are linked to an HP\,8720\,D vector analyzer through flexible cables. All the measurements are performed after a proper calibration to get rid of any parasitic influence of cables and connectors and even of the analyzer itself. The measurements are given in terms of scattering coefficients which form the $S$-matrix $\left(\begin{smallmatrix} S_{11} & S_{12}\\S_{21} & S_{22}\end{smallmatrix}\right)$, where $S_{11}$ (resp.  $S_{22}$) measures the reflection on port 1 (resp. 2) and $S_{12}$ (resp.  $S_{21}$) measures the transmission from port 2 (resp. 1) to port 1 (resp. 2).

In the following section we develop an $S$-matrix formulation for an ideal cavity through the introduction of an electromagnetic model of antennas which enables us to write the response of the cavity within the form of a Breit-Wigner decomposition. Then, in the third section, we complete this description with a perturbative evaluation of ohmic losses at the walls of the cavity. We show that the resulting ohmic width of each resonance may be decomposed in two qualitatively and quantitavely distinct contributions, one of them being sensitive to the spatial distribution of the wavefunction at the contour. Then, in section \ref{exp}, we proceed to an experimental validation of our model in the case of a rectangular cavity. We show that, for each measured resonance, we are able to discriminate quantitatively among the two ohmic contributions to the total widths and the contribution due to the presence of the antennas. 

\section{$S$-matrix formulation for a cavity without ohmic losses}
\subsection{Electromagnetic model of antennas}

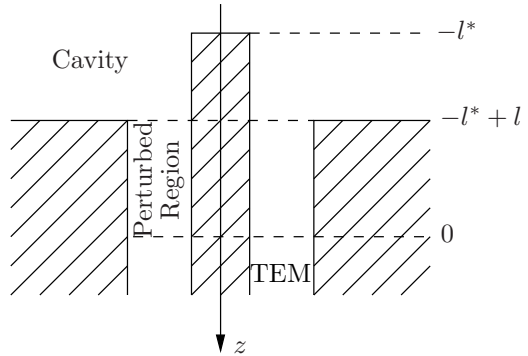
\begin{figure}[h]
\begin{center}
\input{antenne_model.pstex_t}
\end{center}
\caption{Schematic cut view of a coupling antenna. Physical regions introduced in our model are displayed along with the associated characteristic lengths.}
\label{antenna}
\end{figure}

As an antenna, we use  the terminal part of the center conductor of a coax (see Figure \ref{antenna}). Far from this termination,   in a coaxial line, only transverse electro-magnetic (TEM) modes can propagate and the field results as the superposition of incoming and outgoing parts. In the vicinity of the termination of the line, hereafter called the \emph{perturbed region}, perturbative non-propagating waves exist\cite{Harringtonbook}. The longitudinal variable $z$ along the line is oriented outward from the cavity and its origin located at the border between the TEM and the perturbed regions, i.e. at a distance $l^*$ from the end of the antenna (see Figure \ref{antenna}). In the perturbed region, assuming a sinusoidal behavior, we write the stationary current $I_{pert}(z)$  as 
\begin{equation}
\label{I_pert}
I_{pert}(z)=I_+ \er^{\ir k z} + I_- \er^{-\ir k z} \textrm{ for } -l^* < z < 0\,,
\end{equation}
the time evolution being conventionally written $\exp(- \ir \omega t)$.

The current $I(z)$ in the TEM region reads\cite{Collinbook}:
\begin{equation}
\label{I_TEM}
I(z) = \frac{V_0}{Z} (\mathcal{A}_{out}\er^{\ir k z} + \mathcal{A}_{in}\er^{-\ir k z}) \text{ for } z > 0\,,
\end{equation}
where $Z$ is the characteristic impedance of the coaxial line. The following continuity conditions are then imposed
\begin{equation}
\label{limites_I}
\begin{cases} I_{pert}(-l^*)=0 \\  I_{pert}(0)=I(0) \end{cases}\,.
\end{equation}
Therefore
\begin{equation}
\label{I_final}
I_{pert}(z)= I(0)  \frac{\sin k(l^*+z)}{\sin kl^*} \text{ pour } -l^* < z < 0
\end{equation}
with:
\begin{equation}
\label{I0}
 I(0) = \frac{V_0}{Z}(\mathcal{A}_{out}+\mathcal{A}_{in})
\end{equation}

Inside the cavity, time-independent Maxwell's equations yield the following wave equation for the electric field $\vec{E}$:
\begin{equation}
\label{maxwell5}
\triangle \vec{E} + k^2 \vec{E}= \ir \omega \mu \vec{J}\,,
\end{equation}
where, assuming a pointlike antenna at location $\vec{r}_0$, the current density  within the plane of the cavity reads:
\begin{equation}
\label{densite_courant}
\vec{J}(z,\vec{r})= I(z)\delta(\vec{r}-\vec{r}_0)\hat{z} \text{ for } -l^*< z <0\,.
\end{equation}
The 2D formulation of our problem is obtained by integrating equation (\ref{maxwell5}) along $z$ in different ways for the left-hand side and the right-hand side. Indeed, while the left-hand side is easily integrated over the thickness $d$ of the cavity, \textit{i.e.} for $-l^*+l-d < z < -l^*+l$, the integration of the right-hand side is more involved. To account for the effective coupling of the electric field with the current in the perturbed region, we define a coupling function $f(z)$ on the interval  $[-l^*,0]$ which multiplies the current before integrating. To our knowledge, only numerical approaches of this problem have been published: using FDTD\cite{Liu_2000} or modal decomposition\cite{Shen_1996, Eom_2000}. Here we adopt an effective description by assuming $f(z)=1$ on the interval $[-l^*,-l^*+l_{eff}]$ and $f(z)=0$ on the rest of the interval. The length $l_{eff}$ is an adjustable parameter lying between $l$ and $l^*$, most likely close to $l$.
 Hence, for $M$ identical antennas, equation (\ref{maxwell5}) becomes 
\begin{widetext}
\begin{align}
\label{maxwell6} (\triangle  + k^2) E_z(\vec{r})\times d&= \ir \omega \mu_0 \sum_{c=1}^M \delta(\vec{r}-\vec{r}_c) \frac{V_0}{Z} (\mathcal{A}_{out}^c +
\mathcal{A}_{in}^c) \int_{-l*}^{-l^*+l_{eff}} \frac{\sin k(l^*+z)}{\sin kl^*} \dr z \notag\\
&= \frac{\ir V_0 Z_0}{Z}  \frac{\sin^2 \frac{k l_{eff}}{2}}{\sin \frac{k
l^*}{2} \cos \frac{k l^*}{2}} \sum_{c=1}^M \delta(\vec{r}-\vec{r}_c) (\mathcal{A}_{out}^c + \mathcal{A}_{in}^c)
\end{align}
\end{widetext}
where $Z_0= \sqrt{\mu_0 / \epsilon_0}$ is the vacuum impedance. 

\subsection{Breit--Wigner decomposition}

We now follow a standard approach in scattering theory to analytically express the $S$-matrix of a microwave cavity coupled to pointlike antennas. This kind of calculations was initiated in nuclear physics \cite{Feshbachbook} and has been reproduced in various contexts since then (see e.g. references \cite{Dittes_2000}, \cite{Fyodorov_1997} and  \cite{Albeverio_1996}). The cavity is described as a closed system coupled to $M$ channels, one for each antenna. The complete Hilbert space of the system comprises the cavity and the channels. It is therefore decomposed as the direct sum of Hilbert spaces associated to the inside and the outside of the cavity: $\mathcal{E}= \mathcal{E}_{in}\oplus\mathcal{E}_{out}$. Though the inside Hilbert space is of infinite dimension, we adopt the commonly used simplification of a finite dimension $N \gg 1$ \cite{Dittes_2000}. The Hamiltonian of the cavity $\mathcal{H}_{in}$ is thus represented by an $N \times N$ matrix $H$. The eigenstates associated to $\mathcal{H}_{in}$ are denoted  $| \mu \rangle$. The outside Hilbert space is associated to the $M$ antennas and is written as the direct sum: $\mathcal{E}_{out}(E)=\mathcal{E}_1(E) \oplus \dotsi \oplus \mathcal{E}_M(E)$, where $E$ is the energy of  a continuum of scattering states denoted by $| c , E \rangle$ for channel $c$. Finally $W$ denotes the coupling matrix of dimension $N\times M$ between the bound states of the cavity and the scattering states of the antennas. As long as the wavelength remains smaller than the distances between antennas, the direct coupling between channels may be neglected. The complete Hamiltonian $\mathcal{H}$ thus reads:
\begin{widetext}
\begin{equation}
\label{hamiltonien_depart}
\begin{split}
\mathcal{H} &=\sum_{\mu ,\,\nu = 1}^N | \mu \rangle H_{\mu \nu}\langle \nu | 
+ \sum_{c = 1}^M \int \dr E \, | c , E \rangle E \langle c , E | 
+  \sum_{c = 1}^M \sum_{\mu = 1}^N \left( | \mu \rangle \int \dr E \, W_{\mu c}(E)\langle c , E  | 
+ \text{h.c.}\right)\\
&=\mathcal{H}_{in}+ \mathcal{H}_{out} + \mathcal{W}_{out \rightarrow in}
+ \mathcal{W}_{in \rightarrow out}\,.
\end{split}
\end{equation}
\end{widetext}
The normalization conditions are:
\begin{equation}
\label{normalisation_etats_propres}
\langle \nu | \mu \rangle = \delta_{\nu \mu} \: \text{and}\: 
\langle a , E |b , E' \rangle = \delta_{ab}\delta(E-E')\,.
\end{equation}
The space representation of $\mathcal{H}_{in}$ is given by:
\begin{equation}
\label{Hin_spatial}
\langle \vec{r}\, |\mathcal{H}_{in}| \vec{r}\,' \rangle= - \delta(\vec{r}- \vec{r}\,')\Delta_{\vec{r}}\,.
\end{equation}
Likewise, the space representation of $\mathcal{H}_{out}$ in the coax reads:
\begin{equation}
\label{Hout_spatial}
\langle z_a |\mathcal{H}_{out}| z_b \rangle = - \langle z_a | z_b \rangle \frac{\dr^2}{\dr z^2_b}\,.
\end{equation}
For pointlike antennas the coupling will be represented by:
\begin{equation}
\label{couplage_spatial}
\langle \vec{r}\, | \mathcal{W}_{out \rightarrow in}(E)| z_c \rangle=t_c(E,z_c) \; \delta(\vec{r}- \vec{r}_c)\,.
\end{equation}

Let $\boldsymbol{\Phi}= \bigl( \begin{smallmatrix} \varphi_\alpha(\vec{r}) & \varphi_{\beta_1}(z_1)&\hdots&\varphi_{\beta_M}(z_M) \end{smallmatrix}
\bigr)^T$ denote a state of the complete system. Then the eigenvalue problem ${\mathcal{H}}\boldsymbol{\Phi}=E_0 \boldsymbol{\Phi}$ may be written as:
\begin{widetext}
\begin{equation}\label{pb_vp}
\left\{\begin{array}{rcl}
 -\Delta \varphi_\alpha(\vec{r}) + \sum_{c=1}^M  \delta(\vec{r}- \vec{r}_c) \int \dr z_c \, t_c(E_0,z_c)\varphi_{\beta_c}(z_c) &=& E_0 \varphi_\alpha(\vec{r})\\
t_1^*(E_0,z_1)\varphi_\alpha(\vec{r}_1) - \varphi_{\beta_1}''(z_1) &=& E_0 \varphi_{\beta_1}(z_1)\\
&\vdots&\\
t_M^*(E_0,z_M)\varphi_\alpha(\vec{r}_M) - \varphi_{\beta_M}''(z_M) &=& E_0 \varphi_{\beta_M}(z_M)
\end{array} \right.\,,
\end{equation}
\end{widetext}
where the prime symbols stands for the ordinary derivative.

According to the electromagnetic description given at the beginning of this section, $t_c(E_0,z_c)$ will be vanishing except in a perturbed region of length $l^*$ from the termination of the antenna. It is therefore quite natural to fix the origin of $z_c$ at the limit of this region. For $z_c > 0$, the field is written as the superposition of ingoing and outgoing waves 
\begin{equation}
\label{phibeta_pos}
\varphi_{\beta_c}(z_c)=\frac{1}{\sqrt{k}}(\mathcal{A}_{out}^c \er^{\ir k z_c} + \mathcal{A}_{in}^c \er^{-\ir k z_c}) \text{ for } z_c > 0\,,
\end{equation}
where $k^2=E_0$. The factor $1/\sqrt{k}$ is required for dimensional and normalization purposes, and $\mathcal{A}_{in}^c$ and  $\mathcal{A}_{out}^c$ are dimensionless complex amplitudes. For $z_c\in[-l^*,0]$, the perturbed field $\phi(z_c)$ in the coupling region is still to be a superposition of plane waves. Continuity conditions in agreement with our electromagnetic description imply:
\begin{equation}
\label{limites_phi}
\begin{cases} \phi_{\beta_c}(-l^*)=0 \\  \phi_{\beta_c}(0)= \varphi_{\beta_c}(0) \end{cases}\,.
\end{equation}
One therefore deduces
\begin{equation}
\label{phibeta_neg2}
\phi_{\beta_c}(z_c)=\varphi_{\beta_c}(0) \frac{\sin k(z_c+l^*)}{\sin kl^*} \text{ for } -l^*<z_c < 0\,.
\end{equation}
The $M \times M$ $S$-matrix being defined by: 
\begin{equation}
\label{def_matriceS}
\boldsymbol{\mathcal{A}}_{out}= S\boldsymbol{\mathcal{A}}_{in}\,
\end{equation}
where $\boldsymbol{\mathcal{A}}_{in}= \bigl( \begin{smallmatrix}
\mathcal{A}_{in}^1&\hdots&\mathcal{A}_{in}^M \end{smallmatrix} \bigr)^T$ and $\boldsymbol{\mathcal{A}}_{out}= \bigl( \begin{smallmatrix}
\mathcal{A}_{out}^1&\hdots&\mathcal{A}_{out}^M \end{smallmatrix} \bigr)^T$, only the functions $\varphi_{\beta_c}(z_c)$ for $z_c>0$ are relevant. Equation (\ref{pb_vp}) can thus be obviously reduced to:
\begin{widetext}
\begin{equation}
\label{hamiltonien_matrice3}
\left\{\begin{array}{rcl}
-\Delta \varphi_\alpha(\vec{r}) + \sum_{c=1}^M
\delta(\vec{r}- \vec{r}_c) \int_{-l^*}^0 \dr z_c \, t_c(E_0,z_c)\phi_{\beta_c}(z_c) &=& E_0 \varphi_\alpha(\vec{r})\\
- \varphi_{\beta_1}''(z_1) &=& E_0 \varphi_{\beta_1}(z_1)\\
&\vdots&\\
- \varphi_{\beta_M}''(z_M) &=& E_0 \varphi_{\beta_M}(z_M)\,,
\end{array} \right.
\end{equation}
\end{widetext}
where $t_c(E_0,z_c)=0$ has been used for $z_c>0$. Here again we introduce the simplification of a non-vanishing constant value $t_c(E_0)$ for $t_c(E_0,z_c)$ only on the interval $[-l^*,-l^*+l_{eff}]$. With this assumption and expression (\ref{phibeta_neg2}), the first equation in (\ref{hamiltonien_matrice3}) reads:
\begin{equation}
\label{hamiltonien_matrice4}
-\Delta \varphi_\alpha(\vec{r}) + \sum_{c=1}^M T_c(k)\delta(\vec{r}- \vec{r}_c)\varphi_{\beta_c}(0)=E_0\varphi_\alpha(\vec{r})\,, 
\end{equation}
with 
\begin{equation}
\label{Tcdek}
T_c(k)=\frac{t_c(E_0)}{k}  \frac{\sin^2 \frac{k l_{eff}}{2}}{\sin \frac{k l^*}{2} \cos \frac{k l^*}{2}}\,.
\end{equation}
Assuming identical antennas, \textit{i.e.} $T_c(k)\equiv \tilde{T}(k)$, and by identifying (\ref{maxwell6}) with (\ref{hamiltonien_matrice4}), one deduces 
\begin{equation}
\label{analogie3}
\tilde{T}(k) = \ir \sqrt{k}\frac{ Z_0}{ Z}  \frac{\sin^2 \frac{k l_{eff}}{2}}{\sin \frac{k l^*}{2} \cos \frac{k l^*}{2}}\,.
\end{equation}

In spite of the reduction performed in equations (\ref{hamiltonien_matrice3}), by eliminating the source terms associated to the field inside the cavity\cite{Albeverio_1996, Fyodorov_1997}, the condition of self-adjointness can be recovered through appropriate boundary conditions. In appendix \ref{self} we show that it can be done through the following boundary condition for channel $c$:
\begin{equation}
\label{bords}
\tilde{T}^*(k) \varphi_{\alpha}(\vec{r}_c)= \varphi'_{\beta_c}(0)\,.
\end{equation}

We are now in a position to derive an explicit expression for the $S$-matrix. If the energy dependence of the coupling is small on a scale of the order of the mean energy spacing between neighboring modes, the $S$-matrix can be written\cite{Dittes_2000}:
\begin{equation}
\label{matriceS_depart}
S_{ab} = \delta_{ab}- 2 \pi \ir  \langle a , E | W^{\dag} (E - H_{eff})^{-1} W | b , E \rangle\,,
\end{equation}
where
\begin{equation}
\label{Heff} H_{eff}= H - \ir \pi W W^{\dag}\,.
\end{equation}
The $S$-matrix can thus be rewritten:
\begin{equation}
\label{matriceS_simple} S = \frac {1 - \ir K}{1 + \ir K}\,,
\end{equation}
with
\begin{equation}
\label{K_def} K = \pi W^{\dag} \frac{1}{E-H}W\,.
\end{equation}
Standard linear algebra (see e.g. reference \cite{Fyodorov_1997}) transforms expression (\ref{matriceS_simple}) into  
\begin{equation}
\label{matriceS}
S= I_M -2 \ir\pi W^\dag\frac{1}{k^2-H+\ir\pi W W^\dag}W\,.
\end{equation}
Then, assuming a weak coupling, a perturbative expansion to leading order (considering the  isolated cavity as the ``zeroth order") yields the following expression for an element of the $S$-matrix:
\begin{equation}
\label{element_matriceS2}
S_{ab}= \delta_{ab} -2 \ir\pi \sum_{\mu=1}^N\frac {W^*_{\mu a}W_{\nu b}}{k^2-k^2_\mu+ \ir\pi \sum_{c=1}^{M}|W_{\mu c}|^2}\,,
\end{equation}
where the sum runs over the eigenstates of the isolated cavity with energies $k_\mu^2$. In appendix \ref{green} it is shown that the coupling matrix elements, in the space representation, are given by:
\begin{equation}
\label{defW}
W_{\mu c}= \frac{\tilde{T}(k) \psi_\mu^*(\vec{r}_c)}{\sqrt{k \pi}}\,,
\end{equation}
where $\psi_\mu(\vec{r})=\langle\vec{r}\,\vert\mu\rangle$ is the eigenfunction associated to $k_\mu^2$.  One finally obtains the following explicit expression for the $S$-matrix elements:
\begin{equation}
\label{element_matriceS4}
\begin{split}
S_{ab}= \delta_{ab} &- 2 \ir \frac{|\tilde{T}(k)|^2}{k} \times\ldots\\
&\sum_{\mu=1}^N\frac{\psi_\mu(\vec{r}_a)\psi_\mu^*(\vec{r}_b)}
{k^2-k_\mu^2+ \frac{\ir}{k} |\tilde{T}(k)|^2 \sum_{c=1}^{M} |\psi_\mu(\vec{r}_c)|^2}\,.
\end{split}
\end{equation}

\section{Perturbative evaluation of ohmic losses}
\label{perturbloss}
In this section we present the results deduced from a standard first-order
perturbation approach whose validity is restricted to nondegenerate modes
(see Jackson's textbook \cite{Jacksonbook}). 
The power $\dr P$ dissipated (ohmic losses) by a wave with frequency
$\omega$ within the surface element $\dr a$ of a conductor is given by the
flux of the real part of the Poynting vector through this surface.  By
adopting Jackson's convention $\exp(-\ir\omega t)$ for the
time dependence  of the field, one has
\begin{equation}
\label{perte_dep} 
\frac{\dr P}{\dr a} = -\frac{1}{2} \, \Re[\hat{n} \cdot (\vec{E} \wedge
\vec{H}^*)]
\end{equation}
where $\hat{n}$ is the unit normal vector directed toward the interior of
the conductor and $\vec{E}$ and $\vec{H}$ are the fields at the surface. If
the conductor is perfect, $\vec{E}$ is perpendicular to the surface,
$\vec{H}$ is parallel and there is no dissipated power --- in the following,
parallel or perpendicular will be understood \emph{with respect to the
surface of the conductor}. This ideal situation will
correspond to the zeroth order of our description of the field near the
surface of the actual conductor. For a finite conductivity $\sigma$, one can
compute the first order corrections for the fields following the standard
approach described for instance in reference \cite{Jacksonbook}.

To first order, the perpendicular electric field and the parallel magnetic
field outside the conductor remain unmodified. Using appropriate boundary
conditions together with Maxwell equations, it may be shown that
nonvanishing parallel components of both electric and magnetic fields exist
inside the conductor. These fields decrease as $\exp(-z/\delta)$, where
$\delta = \sqrt{2/\mu_c\sigma\omega}$ is the skin depth ($\mu_c$ being the
magnetic permeability of the conductor and $\sigma$ its effective
conductivity), and only depend on the zeroth order parallel component of the magnetic field $\vec{H}_{\sslash}^{(0)}$ at the surface
of the conductor. By continuity, one deduces the existence of a small
parallel component of the electric field just outside the conductor:
\begin{equation}
\label{E_parallele}
\vec{E}_\sslash^{(1)} = \sqrt{\frac{\omega   \mu_c}{2   \sigma}}
(1-\ir)(\hat{n}\wedge \vec{H}_\sslash^{(0)})\,.
\end{equation}
Using \eqref{E_parallele} one finds:
\begin{equation}
\label{perte_H}
\frac{\dr P}{\dr a} = \frac{\mu_c \omega \delta}{4}
\|\vec{H}_{\sslash}^{(0)}\|^2\,.
\end{equation}
To obtain the total power dissipated through ohmic losses within a cavity, a
mere integration of equation \eqref{perte_H} over the walls is required.

In an ideal 2D cavity, the electromagnetic fields does not vary along $z$:
\begin{equation}
\label{champs_cavite}
\vec{H}^{(0)} = \left\{\begin{array}{l} H_x^{(0)}(x,y) \\ H_y^{(0)}(x,y) \\
0 \end{array} \right. \ \textrm{and} \ \vec{E}^{(0)}
=\left\{\begin{array}{l} 0 \\ 0 \\ E_z^{(0)}(x,y) \end{array} \right.\,.
\end{equation}
Denoting $\psi^{(0)} =  E_z^{(0)}(x,y)$, the time-independent Maxwell
equations are reduced to a 2D Helmholtz equation:
\begin{equation}
\label{helm2d}
(\vec{\nabla}_t^2 + \epsilon\mu\omega^2)\psi^{(0)} = 0
\end{equation}
where the transverse gradient operator $\vec{\nabla}_t$ is associated to the
$(x, y)$ coordinates, and $\epsilon$ and $\mu$ are respectively the
permittivity and the permeability inside the cavity. On the contour,
$\psi^{(0)}$ obeys Dirichlet conditions:
\begin{equation}
\label{dirichlet}
\psi^{(0)} = 0 \ \textrm{on}\  \mathcal{C}\,.
\end{equation}
This yields a complete
analogy with the free propagation of a quantum particle in a 2D infinite well. This
type of system is commonly called a \emph{quantum billiard}.
For a given mode, the ohmic dissipated power reads:
\begin{multline}
\label{perte_int}
P = \frac{1}{2 \sigma \delta}
\Biggl[\oint_\mathcal{C} \dr \ell \int\limits_0^d\dr z\,\|\hat{n} \wedge
\vec{H}^{(0)}\|^2_{\textrm{cont}}\\
+\iint_\mathcal{S}\dr a\,\|\hat{n} \wedge
\vec{H}^{(0)}\|^2_{\textrm{ends}}\Biggr]\,.
\end{multline}
One defines the
integrals $I_1$ and $I_2$ so that equation \eqref{perte_int} is re-written:
\begin{equation}
\label{perte_int2}
P = \frac{I_1+I_2}{2 \sigma \delta \mu}\,.
\end{equation}
Using Faraday's law, one obtains:
\begin{equation}
\label{I1_final}
I_1 =\xi\frac{Ld}{A}\epsilon\iint_\mathcal{S}\dr a\,|\psi^{(0)}|^2
\end{equation}
where $\xi$ is defined by:
\begin{equation}
\label{defxi}
\frac{1}{\epsilon\mu\omega^2} \oint_\mathcal{C}\dr
\ell\,|\partial_n\psi^{(0)} |^2 \equiv \xi\frac{L}{A}\iint_\mathcal{S}\dr
a\,|\psi^{(0)} |^2\, ,
\end{equation}
$\partial_n$ denoting $(\hat{n} \cdot \vec{\nabla})$. Here it should be
remarked that $\xi$ is a parameter, which depends on the spatial structure
of the mode at hand. Likewise, one may compute $I_2$:
\begin{equation}
\label{I2}
I_2 = \frac{2}{\mu\omega^2} \iint_\mathcal{S}\dr a\,\|\vec{\nabla}_t
\psi^{(0)}\|^2\,.
\end{equation}
Now using the 2D Green's theorem together with equation \eqref{helm2d}, one
gets:
\begin{equation}
\label{I2_final}
I_2 = 2\epsilon\iint_\mathcal{S}\dr a\,|\psi^{(0)}|^2
\end{equation}
and, eventually, 
\begin{equation}
\label{perte_int3}
P = \frac{\epsilon}{\sigma \delta\mu}
\left(1+\xi \frac{Ld}{2A}\right)\iint_\mathcal{S}\dr a\,|\psi^{(0)}|^2\,.
\end{equation}
Considering that the total electromagnetic energy stored in the cavity is
given by:
\begin{equation}
\label{W}
W = \frac{\epsilon}{2} \iiint_\mathcal{V} \dr v \,\|\vec{E}^{(0)}\|^2
= \frac{\epsilon d}{2} \iint_\mathcal{S}\dr a\,|\psi^{(0)}|^2\,,
\end{equation}
the FWHM of the resonances is given by
\begin{equation}
\label{gamma}
\Gamma = \frac{P}{W}
= \frac{\mu_c}{\mu}\frac{1}{d} \sqrt{\frac{2\omega}{\mu_c\sigma}}\left(1+\xi
\frac{Ld}{2A}\right)\,.
\end{equation}
In our context $\mu_c/\mu$ is practically equal to unity.
Thus one finally has to consider two distinct types of ohmic losses: those
located at the surface of both ends, which amount to attenuation along
propagation, and ohmic losses upon reflection at the contour:
\begin{equation}
\label{gamma_ohm}
\Gamma^{\textrm{ohm}} = \Gamma^{\textrm{prop}}+\Gamma^{\textrm{refl}}
\end{equation}
where
\begin{equation}
\label{gamma_prop}
\Gamma^{\textrm{prop}}= \frac{2}{d} \sqrt{\frac{\omega}{2 \mu
\sigma_{ends}}}= \frac{\delta_{\textrm{ends}} \omega}{d}\, ,
\end{equation}
\begin{equation}
\label{gamma_bords}
\Gamma^{\textrm{refl}}= \xi \frac{L}{A} \sqrt{\frac{\omega}{\pi \mu
\sigma_{\textrm{cont}}}}= \xi \frac{L \delta_{\textrm{cont}} \omega}{2
A}\, .
\end{equation}
Here, we have introduced two different effective conductivities
($\sigma_{\textrm{ends}}$ and $\sigma_{\textrm{cont}}$) and their
corresponding skin depths ($\delta_{\textrm{ends}}$ and
$\delta_{\textrm{cont}}$) to account for two different types of copper
used in our experiment, and for possible different surface states for the top
and bottom plates and the inner surface of the copper frame used as the
contour.
These two contributions to the widths are also quite distinct in their
physical interpretation. Indeed, $\Gamma^{\textrm{prop}}$ truly corresponds
to losses endured by a plane wave propagating in free space between two
parallel infinite metallic planes. It is a slowly varying function of
frequency, depending neither on the transverse geometry of the cavity, nor
on the spatial distribution of the wavefunction in the cavity.
$\Gamma^{\textrm{refl}}$, to the opposite, is related to a loss mechanism
located on the contour of the cavity, which clearly depends on the geometry
of the latter, and chiefly, on the spatial distribution of the normal
gradient of the wavefunction via the quantity $\xi$.
$\Gamma^{\textrm{refl}}$ therefore fluctuates from mode to mode and, in the case of the rectangular cavity, its explicit form will be given in the following section.
By using a boundary perturbation technique to compute losses pertaining to
reflections on the contour it is shown in appendix \ref{perturb} that, to each
correction of the imaginary part of the frequency due to ohmic losses, there
is a correspondingly equal correction of the real part. Moreover this
boundary approach sheds light on the intimate connection between the
boundary conditions on the contour, leading to non-purely real
wavefunctions, and the fluctuating partial widths
$\Gamma^{\textrm{refl}}$\cite{qfactor}.

Collecting the above results with those obtained in the previous section we
are now in a position to write the $S$-matrix between weakly coupled
pointlike antennas in a 2D cavity in the presence of ohmic losses. It reads:
\begin{equation}
\label{matriceS_final}
S_{ab}= \delta_{ab} - 2 \ir T^2(\omega)
\sum_{n=1}^N \frac{\psi_n(\vec{r}_a) \psi_n(\vec{r}_b)}
{\omega^2-\omega_n^2+\ir \omega^{(0)}_n
(\Gamma_n^{\textrm{ohm}}+\Gamma_n^\textrm{ant})}
\end{equation}
where $\omega_n=\omega_n^{(0)}-\Gamma_n^{\textrm{ohm}}/2$ and the $\{\omega_n^{(0)}\}$'s are
the unperturbed eigenfrequencies of the ideal lossless cavity, and where
\begin{equation}
\label{Tdef}
 T(\omega) = c \frac{Z_0}{Z}  \frac{\sin^2 \frac{\omega
l_{eff}}{2c}}{\sin \frac{\omega l^*}{2c} \cos \frac{\omega l^*}{2c}}\,,
\end{equation}
the contribution of antennas to the widths being given, at leading order, by
\begin{equation}
\label{gamma_ant}
\Gamma_n^\textrm{ant} = \frac{T^2(\omega_n)}{\omega^{(0)}_n}
\sum_{c=1}^{M} |\psi^{(0)}_n(\vec{r}_c)|^2 \,.
\end{equation}
Here  it should be remarked that the second factor $\psi$ in equation
(\ref{matriceS_final}) should not be a complex conjugate. Indeed, due to ohmic losses, the wavefunctions are no longer real and the $S$-matrix cannot  keep its unitarity. Nonetheless, it obviously has to remain symmetric. In the previous section we used the Hermitian formalism for the sake of convenience, but it turns out to be inappropriate for the present purpose (see for instance reference \cite{Collinbook} about self-adjoint systems).

\section{Experimental validation in a rectangular cavity}
\label{exp}

\subsection{A preliminary global test}
\begin{figure}[h]
\input{test_transmission.pstex_t}
\caption{Experimentally based estimation of $T^2(\omega_n)$ as given by (\ref{S12_moy}) (crosses). The gray region corresponds to its theoretical expression (\ref{Tdef}) for values of $l_{eff}$ between 1.7\,mm and 2.3\,mm ($l^*=1.65$\,cm).}
\label{transmission_brut}
\end{figure}
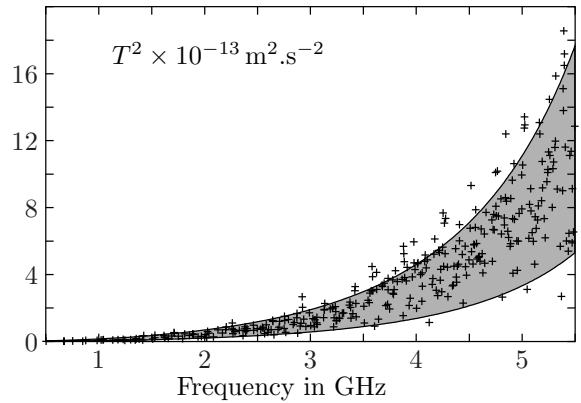

The aim of this section is to check the pertinence of the description given above in a rectangular cavity where eigenfunctions and eigenfrequencies are easily calculated in the limit of vanishing losses. A preliminary test, which does not involve any sophisticated fitting procedure, consists in comparing the average transmission between two antennas with the corresponding quantity deduced from equation (\ref{matriceS_final}). According to this equation, the transmission coefficient ($a \neq b$) for $\omega=\omega_n$ approximately reads:
\begin{equation}
\label{S12_final}
 |S_{ab}(\omega_n)|\simeq 2 T^2(\omega_n)
\frac{|\psi_n(\vec{r}_a) \psi_n(\vec{r}_b)|} { \omega^{(0)}_n
\Gamma_n}\,,
\end{equation}
where $\Gamma_n=\Gamma_n^{\textrm{ohm}}+\Gamma_n^\textrm{ant}$. In a lossless rectangular cavity with sides $L_x \times L_y$, the eigenfrequencies are:
\begin{equation}
\label{freq_rectangle}
\omega_{l,m}^{(0)}= \pi c \sqrt{\left(\frac{l}{L_x}\right)^2+\left(\frac{m}{L_y}\right)^2}\,,
\end{equation}
where $l,m$ are integers. The corresponding eigenfunctions read:
\begin{equation}
\label{modes_rectangle} \psi_{l,m}^{(0)}(x,y)=\frac{2}{\sqrt{L_xL_y}} \sin \frac{ l
\pi x}{L_x}  \sin \frac{ m \pi y}{L_y}\,.
\end{equation}
Assuming that $|\psi_n(\vec{r})|\simeq|\psi_{l,m}^{(0)}(x,y)|$, one deduces:
\begin{equation}
\label{S12_moy}
T^2(\omega_n) \simeq \frac{L_xL_y \pi^4 \omega_n \Gamma_n}{128} \langle|S_{ab}(\omega_n)|\rangle \,,
\end{equation}
where the average is performed on the positions $\vec{r}_a$ and $\vec{r}_b$ of the antennas. In our experiments the average was obtained by performing the 45 distinct transmission measurements that the 10 antennas allow. The values of $\omega_n$ and $\Gamma_n$ used in the experimental evaluation of the right-hand side of (\ref{S12_moy}) were obtained, in a rough way, through the analysis of the \emph{group delay}. This quantity, defined as the derivative of the phase $\varphi_{ab}$ of $S_{ab}$ with respect to $\omega$, presents rather well defined extrema at frequencies close to eigenfrequencies. For a gross estimation of $\Gamma_n$ we used $\Gamma_n\simeq 2/\langle\dr\varphi_{ab}(\omega_n)/\dr\omega\rangle$, which is exact for an isolated resonance. Note that, in the case of moderate or large modal overlap, this method generally overestimates the widths. Knowing the dimensions $L_x$ and $L_y$ of the cavity, the right-hand side of (\ref{S12_moy}) only depends on the length parameters $l_{eff}$ and $l^*$. Recall that $l^*$ is the length of the perturbed region at the end of the coax, and that $l_{eff}$ is the effective length over which the field inside the cavity is coupled to the antenna. One may assume, in this preliminary global test, that $l_{eff}$ remains close to the length $l$ of the part of the antenna which lies inside the cavity: $l_{eff}\simeq l=2.0 \pm 0.3$\,mm. In the same way, a rough estimate of $l^*$ is given by the distance between the end of the antenna and the reference (calibration) plane of the coax: $16.5 \pm 0.2$\,mm. Figure \ref{transmission_brut} compares $T^2$ as obtained through equation (\ref{Tdef}) with its experimental value deduced from equation (\ref{S12_moy}) for the first 348 resonances up to 5.5\,GHz. The lowest and the highest continuous curves correspond to $l_{eff}=1.7\,$mm and $l_{eff}=2.3\,$mm respectively. At the resolution of the presented figure, the curves obtained for values of $l^*$ ranging from $16.3\,$mm to $16.7\,$mm are not distinguishable. A fair agreement is observed between the experiment and our model. A first consequence of this test is to substantiate the correspondence between the length parameters of our model and the actual lengths of the coax antennas. Note also that this global preliminary test requires no sophisticated data processing of the individual resonances. In the following, we will see how our model for coupling with antennas remains quite satisfactory when put to more stringent tests.

\subsection{A test with individual resonances}
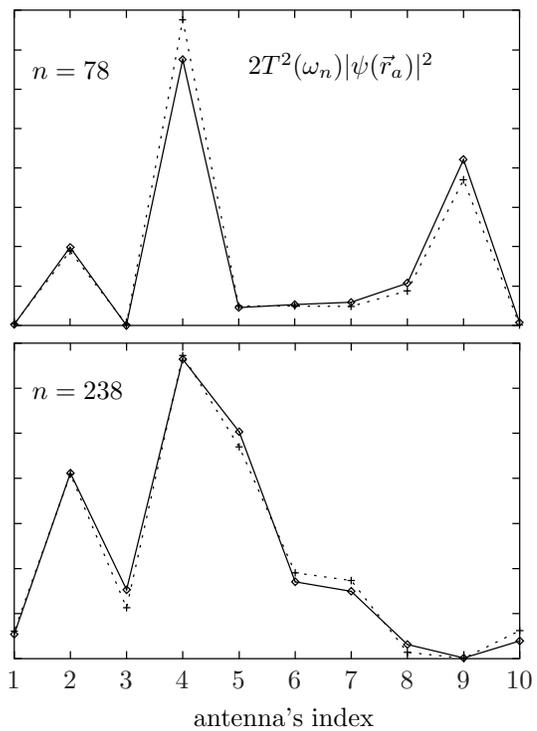
\begin{figure}[h]
\begin{center}
\input{fit_trans_pre.pstex_t}
\end{center}
\caption{Comparison of (\ref{amplitude_moy}) with $2 T^2(\omega_n)|\psi_{l,m}^{(0)}(x_a,y_a)|^2$ at the ten antennas (indices are introduced in figure \ref{cavity}) for resonances $n=78$ and $n=238$.}
\label{compare}
\end{figure}

In the global test presented above we had no need of a precise knowledge of the wavefunctions at the antennas. To check the validity of formula (\ref{matriceS_final}), we have developed an original fitting procedure (see \cite{Jerome_2003}) to extract the actual complex eigenfrequencies $\omega_n$ and the complex amplitudes
\begin{equation}
\label{amplitude}
A_n^{ab}=-2\ir T^2(\omega_n)\psi_n(\vec{r}_a)\psi_n(\vec{r}_b)\,.
\end{equation}
With this procedure, based upon a mixture of robust algorithms, we could check that the actual eigenfrequencies remain very close to the unperturbed values given by (\ref{freq_rectangle}) in the frequency range studied here. Now we proceed to verify that the amplitudes $A_n$ deduced from our measurements are well described by (\ref{amplitude}) with expression (\ref{Tdef}) for $T$ and formula (\ref{modes_rectangle})  to approximate the true eigenfunctions. Indeed, even if the existence of non-uniform losses (chiefly those associated to $\Gamma^{\textrm{refl}}$) leads to non-purely real wavefunctions, we will see below that the corrections remain very small. From (\ref{amplitude}) it is easily deduced for three different antennas $a,b$ and $c$:
\begin{equation}
\label{amplitude_isole_mod}
\frac{|A_{n}^{ab}||A_{n}^{ac}|}{|A_{n}^{bc}|}= 2 T^2(\omega_n)|\psi_n(\vec{r}_a)|^2\,.
\end{equation}
As there are 36 different ways of combining the 45 amplitudes $A^{ab}$, yielding 36 slightly different values of (\ref{amplitude_isole_mod}), one can use the following average estimate:
\begin{equation}
\label{amplitude_moy}
2 T^2(\omega_n) |\psi_{l,m}^{(0)}(x_a,y_a)|^2= \frac{1}{36} \sum_{\substack{b,c \neq a\\ b<c}} \frac{|A_{n}^{ab}||A_{n}^{ac}|}{|A_{n}^{bc}|}\,,
\end{equation}
where $|\psi_n(\vec{r})|\simeq|\psi_{l,m}^{(0)}(x,y)|$ is assumed. For a given resonance one can directly compare the above quantity with $2 T^2(\omega_n)|\psi_{l,m}^{(0)}(x_a,y_a)|^2$ for the ten antennas ($a=1,\ldots,10$). This comparison is shown on figure \ref{compare} for two distinct resonances, namely $n=78$ at 2.655\,GHz and $n=238$ at 4.558\,GHz. As the great majority of resonances that we are concerned with are narrow enough to ensure a good correspondence with the unperturbed wavefunctions, we observe a fairly good agreement. 

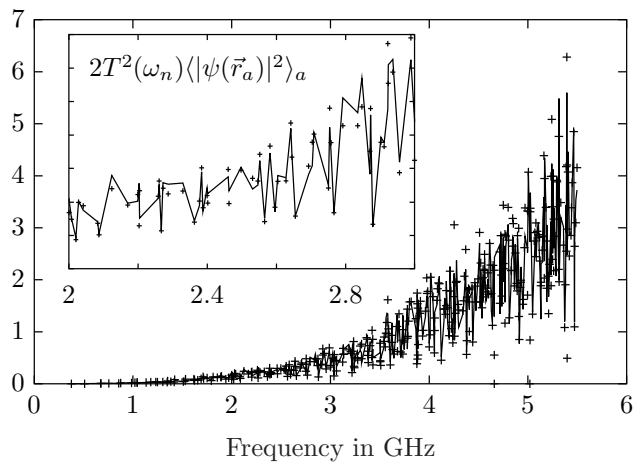
\begin{figure}[h]
\input{fit_trans2_lin_fin.pstex_t}
\caption{Comparison between the experimental quantity $\langle\frac{1}{36} \sum_{b<c\neq a} |A_{n}^{ab}||A_{n}^{ac}|/|A_{n}^{bc}|\rangle_a$ (crosses) with the prediction $2 T^2(\omega_n) \langle|\psi_{l,m}^{(0)}(x_a,y_a)|^2\rangle_a$ (line). Inset : enlarged view within the range 2-3\,GHz.}
\label{compare2}
\end{figure}

To extend our comparison to all resonances, one can now average over all ten locations of antenna $a$. For the best values of $l_{eff}=1.9\,$mm and $l^*=16.5\,$mm obtained in the frequency range from 2\,GHz to 5.5\,GHz, figure \ref{compare2} shows the comparison between the experimental quantity $\langle\frac{1}{36} \sum_{b<c\neq a} |A_{n}^{ab}||A_{n}^{ac}|/|A_{n}^{bc}|\rangle_a$ and the prediction $2 T^2(\omega_n) \langle|\psi_{l,m}^{(0)}(x_a,y_a)|^2\rangle_a$ up to 5.5\,GHz. The agreement is excellent on the average. By a close inspection, e.g. between 2\,GHz and 3\,GHz as shown in the inset, one can notice that the agreement is generally excellent even at the level of individual resonances. Rare important discrepancies are observed for very close neighboring eigenfrequencies (quasi-degeneracies) due to modal overlapping. Indeed, when the latter effect is not negligible, one expects that the spatial distribution of the wavefunction results from a linear combination of  neighboring unperturbed wavefunctions\cite{Kuhl_2000}. Beyond 3\,GHz this effect deteriorates the agreement due to the concomitant increase of the total width and decrease of the mean spacing leading to an inadequacy of the zeroth order eigenfunctions we use for our test.

\subsection{Partial width decomposition}
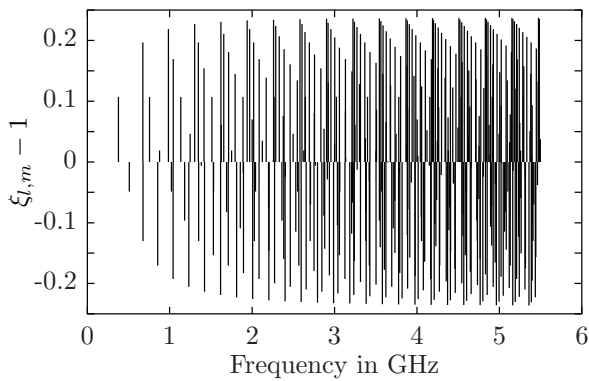
\begin{figure}[h]
\input{xi_impulse.pstex_t}
\caption{Values of $(\xi_{l,m}-1)$ after equation (\ref{xi_lm}) for all resonances up to 5.5\,GHz.}
\label{xilm}
\end{figure}

As seen above, the total width of a resonance can be decomposed as a sum of three partial widths associated to losses through the antennas, ohmic losses on the contour of the cavity, and ohmic losses at the surface of both ends which appear as damping along propagation. In a rectangular cavity, the total width of the $n$th resonance, characterized by the quantum numbers $l$ and $m$, is given by
\begin{equation}
\Gamma_n = \Gamma^{\textrm{ant}}_{l,m}+\Gamma^{\textrm{refl}}_{l,m}
+\Gamma^{\textrm{prop}}(\omega_n)\,.
\end{equation}
By using expression (\ref{modes_rectangle}) for the wavefunctions in order to evaluate the factor $\xi$ in equation (\ref{defxi}), one obtains
\begin{equation}
\label{xi_lm}
\xi_{l,m}= \frac{A c^2}{\omega_{l,m}^2 L} \left(\frac{l^2}{L_x^3}+\frac{m^2}{L_y^3}\right)
\end{equation}
whence, using (\ref{gamma_bords}),
\begin{equation}
\label{gamma_bords_lm_bis}
\Gamma_{l,m}^{\textrm{refl}} \simeq \frac{c^2\delta_{\textrm{cont}}(\omega_n)}{2\omega_{l,m}^{(0)}} \left(\frac{l^2}{L_x^3}+\frac{m^2}{L_y^3}\right)\,.
\end{equation}
These partial widths clearly vary from mode to mode as illustrated in figure \ref{xilm} where $\xi_{l,m}-1$ is shown for each eigenfrequency up to 5.5\,GHz. Note that the $\xi_{l,m}$'s oscillate around unity and vary at most by 23\,$\%$.

The widths associated to losses through antennas also vary from mode to mode:
\begin{equation}
\label{gamma_ant_lm}
 \Gamma_{l,m}^{\textrm{ant}}= \frac{T^2(\omega_n)}{\omega_{l,m}^{(0)}} \sum_{c=1}^{M} |\psi_{l,m}^{(0)}(\vec{r}_c)|^2\,.
 \end{equation}
In the previous subsection we already checked the validity of the above formula for all resonances shown on figure \ref{compare2} since $\Gamma_{l,m}^{\textrm{ant}}$ is essentially proportional to the sum of expression (\ref{amplitude_moy}) over all antennas.

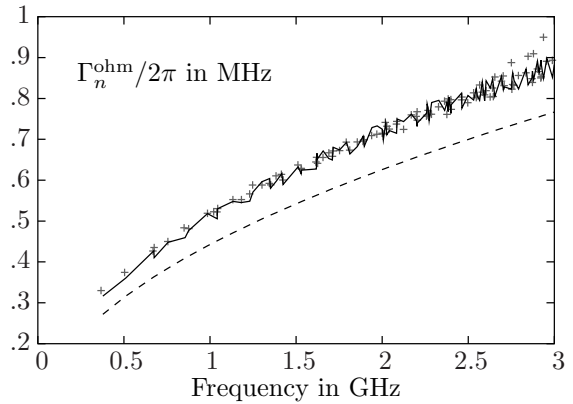
\begin{figure}[h]
\input{largeurs_vide_ins.pstex_t}
\caption{Behavior of the ohmic widths up to 3\,GHz. Comparison is shown between theoretical ohmic widths (crosses) and experimental ohmic widths (continuous curve). The dashed curve indicates the smooth contribution $\Gamma^{\textrm{prop}}(\omega)$.}
\label{largeurs}
\end{figure}

Thus by fitting the experimental transmission by formula (\ref{matriceS_final}), one obtains a direct measure of the total width and an indirect measure of $\Gamma_{n}^{\textrm{ant}}$, thus enabling us to evaluate the two effective conductivities $\sigma_{\textrm{ends}}$ and $\sigma_{\textrm{cont}}$. A representation of all the ohmic widths $\Gamma_n^{\textrm{ohm}}=\Gamma_n -\Gamma^{\textrm{ant}}_{l,m}$ up to 3\,GHz is given in figure \ref{largeurs}. A comparison is shown between theoretical ohmic widths (crosses) and experimental ohmic widths (continuous curve). The smooth dominant contribution given by  $\Gamma^{\textrm{prop}}(\omega)$ is indicated by the dashed curve. The agreement on the mean level and on the amplitude of fluctuations (only present in $\Gamma_{n}^{\textrm{refl}}$) is fairly good whereas, resonance by resonance, the agreement  is not systematic, essentially due to the effect of modal overlap mentioned above. 

\section{Conclusion}
In conclusion, we have tried to provide a better understanding of the physical mechanisms at the origin of resonance broadening in microwave cavities. We have explicitly developed an $S$-matrix model including the frequency dependent coupling of the antennas and accounting for ohmic absorption at the boundaries of a two-dimensional cavity. We have especially emphasized the necessity to distinguish between ohmic attenuation along propagation,  leading to a smooth frequency dependent contribution to the total width, and localized absorption at the contour of the cavity, yielding  a contribution varying from mode to mode. We have performed experiments where we analyzed the transmission versus frequency in terms of a Breit-Wigner decomposition deduced from our model. In the rectangular cavity we used, all the quantities involved in our theoretical description could easily be calculated in the perturbative limit of small or moderate modal overlap. We therefore have been able to validate our approach and provide a very precise estimation of the various contribution to the total widths. In particular, the varying contribution of ohmic losses at the contour could be quantitatively checked at the level of individual resonances, except for quasi-degenerate modes. This approach has recently enabled us to relate the losses at the contour, in a chaotic cavity, to the imaginary part of the wavefunction\cite{qfactor}.

We believe that our present approach is an important step to test existing
or yet to come theories of open or absorptive chaotic wave systems. Indeed, such
theories generally assume that losses are associated to distinct well identified coupling channels\cite{Dittes_2000}. It is nonetheless not obvious that these can be used to describe different sources of loss as damping along propagation or ohmic dissipation at the contour. For instance, absorbing boundaries may be viewed as a number (of the order of $L/\lambda$) of distributed coupling channels. Indeed, to mimic absorption, recent theoretical predictions have been proposed but only in the asymptotic limit of a large number of effective channels with vanishing coupling\cite{Mendez_2003,Savin_2003} (see also \cite{Rozhkov_2003}).

\acknowledgments{The authors gratefully acknowledge helpful discussions
with P. Sebbah, R. L. Weaver and H.-J. St{\"o}ckmann.}

\appendix
\section{Evaluation of the $S$-matrix}
\subsection{Self-adjointness condition}
\label{self}

In the Hilbert space of the complete problem (cavity and antennas), one defines the following scalar product:
\begin{widetext}
\begin{equation}
\label{produit_scalaire}
(\boldsymbol{\Phi},\boldsymbol{\Phi}')= 
\dint_\mathcal{A} \dr \vec{r} \,
\varphi_{\alpha}^*(\vec{r})\varphi_{\alpha'}(\vec{r})
+ \sum_{c=1}^M \int_0^\infty \dr z_c \, \varphi_{\beta_c}^*(z_c) \varphi_{\beta_c'}(z_c)\,.
\end{equation}
With this scalar product, the self-adjointness condition, 
\begin{equation}
\label{autoadjonction} \bigl( {\mathcal{H}}\boldsymbol{\Phi}, \boldsymbol{\Phi}' \bigr) = \bigl( \boldsymbol{\Phi}, {\mathcal{H}}\boldsymbol{\Phi}' \bigr)\,,
\end{equation}
reads

\begin{equation}
\label{autoadjonction3}
\begin{split}
\bigl({\mathcal{H}}&\boldsymbol{\Phi}, \boldsymbol{\Phi}' \bigr)
-\bigl( \boldsymbol{\Phi}, {\mathcal{H}}\boldsymbol{\Phi}' \bigr) 
=\dint_\mathcal{A} \dr \vec{r} \, \bigl[\varphi_{\alpha}^*(\vec{r})
\bigl(\Delta \varphi_{\alpha'}(\vec{r})\bigr)
-\bigl(\Delta\varphi_\alpha(\vec{r})\bigr)^*\varphi_{\alpha'}(\vec{r})\bigr]+\ldots\\
&+\sum_{c=1}^M \bigl(T_c^*(k)
\varphi_{\beta_c}^*(0)\varphi_{\alpha'}(\vec{r}_c)-\varphi_{\alpha}^*(\vec{r}_c)T_c(k) \varphi_{\beta_c'}(0)\bigr)-\sum_{c=1}^M\int_0^\infty
\dr z_c \,\bigl[\varphi_{\beta_c}''^*(z_c)\varphi_{\beta_{c}'}(z_c)-\varphi_{\beta_{c}}^*(z_c)\varphi_{\beta_c'}''(z_c)\bigr]\,.\end{split}
\end{equation}
Using Green's theorem for the first term and integrating the third one by parts, one obtains:
\begin{equation}
\label{autoadjonction4}
\begin{split} \bigl({\mathcal{H}}&\boldsymbol{\Phi}, \boldsymbol{\Phi}' \bigr)-\bigl( \boldsymbol{\Phi},
{\mathcal{H}}\boldsymbol{\Phi}' \bigr) =\oint_\mathcal{C} \dr \hat{n} \bigl[\varphi_{\alpha}(\vec{r})\bigl(\nabla
\varphi_{\alpha'}(\vec{r})\bigr)-\bigl(\nabla \varphi_\alpha(\vec{r})\bigr)\varphi_{\alpha'}(\vec{r})\bigr]+\ldots\\&+\sum_{c=1}^M \bigl(T_c^*(k)
\varphi_{\beta_c}^*(0)\varphi_{\alpha'}(\vec{r}_c)-\varphi_{\alpha}^*(\vec{r}_c)T_c(k) \varphi_{\beta_c'}(0)\bigr)-\sum_{c=1}^M
\bigl[\varphi_{\beta_c}'^*(z_c)\varphi_{\beta_{c}'}(z_c)-\varphi_{\beta_{c}}^*(z_c)\varphi_{\beta_c'}'(z_c)\bigr]_0^\infty\,.\end{split}
\end{equation}
\end{widetext}
The eigenfunctions  $\varphi_\alpha(\vec{r})$ obey boundary Dirichlet conditions, and the last term in (\ref{autoadjonction4}) vanishes for $z\to\infty$. The self-adjointness condition can thus be written:
\begin{equation}
\label{autoadjonction7}
\begin{split}
&\sum_{c=1}^M\bigl[ \bigl( \varphi_{\beta_c}'^*(0)\varphi_{\beta_{c}'}(0)- \varphi_{\beta_{c}}^*(0)
 \varphi_{\beta_c'}'(0) \bigr)+\ldots\\
&+  \varphi_{\beta_c}^*(0) T_c^*(k) \varphi_{\alpha'}(\vec{r}_c)
- T_c(k) \varphi_{\alpha}^*(\vec{r}_c) \varphi_{\beta_c'}(0)\bigr]=0\,.
\end{split}
\end{equation}
This condition is non-trivially fulfilled by imposing:
\begin{equation}
\label{condition_bords}
T_c^*(k) \varphi_{\alpha}(\vec{r}_c)= \varphi'_{\beta_c}(0)\,.
\end{equation}

\subsection{The coupling matrix elements}
\label{green}

Assuming $E_0=k^2$, equation (\ref{hamiltonien_matrice4}) now reads:
\begin{equation}
\label{premiere_ligne}
(\Delta+k^2) \varphi_\alpha(\vec{r})=\sum_{c=1}^M T_c(k) \delta(\vec{r}- \vec{r}_c) \varphi_{\beta_c}(0)\,.
\end{equation}
Furthermore, the Green's functions of the isolated cavity are given by:
\begin{equation}
\label{def_green}
(\Delta + k^2) G(\vec{r}, \vec{r}^{\,\prime}, k^2)= \delta(\vec{r}-\vec{r}^{\,\prime})\,.
\end{equation}
By comparing equations (\ref{premiere_ligne}) and (\ref{def_green}), for $\vec{r} = \vec{r}_b$, one obtains:
\begin{equation}
\label{premiere_ligne2}
\varphi_\alpha(\vec{r}_b)=\sum_{c=1}^M T_c(k)  G(\vec{r}_b, \vec{r}_c, k)\varphi_{\beta_c}(0)\,,
\end{equation}
With the boundary conditions (\ref{condition_bords}) and the expression  (\ref{phibeta_pos}) of $\varphi_{\beta}$, relation (\ref{premiere_ligne2}) becomes:
\begin{widetext}
\begin{equation}
\label{calcul_matriceS3}
\mathcal{A}_{in}^c -\frac{\ir}{k} T_b^*(k) \sum_{c=1}^M T_c(k)  G(\vec{r}_b, \vec{r}_c, k) \mathcal{A}_{in}^b= \mathcal{A}_{out}^c  +\frac{\ir}{k} T_b^*(k) \sum_{b=1}^M T_c(k) G(\vec{r}_b, \vec{r}_c, k) \mathcal{A}_{out}^b\,,
\end{equation}
\end{widetext}
or, in a matrix form,
\begin{equation}
\label{calcul_matriceS4}
\boldsymbol{\mathcal{A}}_{in} -\ir K \boldsymbol{\mathcal{A}}_{in}
= \boldsymbol{\mathcal{A}}_{out} +\ir K  \boldsymbol{\mathcal{A}}_{out}\,,
\end{equation}
where $K$ is an $M \times M$ matrix defined by:
\begin{equation}
\label{defK1}
K_{ab}(k)=  \frac{T_a^*(k)}{\sqrt{k}}G(\vec{r}_a, \vec{r}_b, k)  \frac{T_b(k)}{\sqrt{k}}\,.
\end{equation}
From $\boldsymbol{\mathcal{A}}_{out}= S\boldsymbol{\mathcal{A}}_{in}$ one thus makes the relation (\ref{matriceS_simple}) between $K$ and the $S$-matrix explicit. By introducing in (\ref{defK1}) the expansion of the Green's function in terms of the eigenfunctions $\psi_\nu(\vec{r})=\langle r | \nu \rangle$ of the isolated cavity:
\begin{equation}
\label{green_developpee}
G(\vec{r}, \vec{r}\,', E)
=\sum_{\nu=1}^N\frac{\psi_\nu(\vec{r})\psi_\nu^*(\vec{r}\,')}{k^2-k^2_\nu}\,,
\end{equation}
one finds the following expression for the $K$-matrix
\begin{equation}
\label{defK2} K_{ab}(k)
=\left( \frac{ T_a(k) \psi_\nu^*(\vec{r}_a)}{\sqrt{k}} \right)^* \frac{1}{k^2-k^2_\nu} \left( \frac{T_b(k)
\psi_\nu^*(\vec{r}_b)}{\sqrt{k}}  \right)\,.
\end{equation}
Finally, the elements of the coupling matrix $W$, related to $K$ by equation (\ref{K_def}), are given by:
\begin{equation}
W_{\mu c}= \frac{T_c(k) \psi_\mu^*(\vec{r}_c)}{\sqrt{k \pi}}\,.
\end{equation}

\section{Perturbative boundary conditions}
\label{perturb}

An alternative way of computing the losses at the contour is easily
obtained by following a boundary perturbation technique (see for instance
\cite{Jacksonbook}). Indeed, calling $\psi_0$ the solution at the real
eigenfrequency $\omega_0$ of the zeroth-order problem defined by equations
(\ref{helm2d}) and (\ref{dirichlet}), the perturbation of the boundary
conditions can be written:
\begin{equation}
\label{mixedbc}
\psi \simeq -(1+i) \frac{\mu_c \delta}{2 \mu} \partial_n\psi_0 \
\textrm{on}\  \mathcal{C}\,.
\end{equation}
Green's theorem applied to this 2D problem straightforwardly  yields:
\begin{equation}
\label{eigenpert}
\omega^2 - \omega_0^2  \simeq - (1+i) \frac{1}{\epsilon\mu} \frac{\mu_c
\delta}{2 \mu} \frac{\oint_\mathcal{C}\dr \ell\,|\partial_n\psi_0
|^2}{\iint_\mathcal{S}\dr a\,|\psi_0 |^2} \,.
\end{equation}
Writing the perturbed eigenfrequency as $\omega = \omega_0 + \delta\omega -
i\Gamma/2$, equation (\ref{eigenpert}) becomes:
\begin{equation}
\label{eigenpert2}
\delta\omega - i\Gamma/2  \simeq - (1+i) \frac{1}{\epsilon\mu} \frac{\mu_c
\delta}{4 \mu \omega_0} \frac{\oint_\mathcal{C}\dr \ell\,|\partial_n\psi_0
|^2}{\iint_\mathcal{S}\dr a\,|\psi_0 |^2} \,,
\end{equation}
leading, in the case at hand, to equal corrections on both real and
imaginary parts of $\omega$. This last comment is general and applies as
well to the corrections originating from ohmic losses at top and bottom of
the cavity. Thus, this perturbative approach completely agrees with the
one developed in section \ref{perturbloss} as long as the widths are concerned and completes it as it
provides an estimation of the frequency shift of the resonances related to
ohmic losses.  It is also particularly interesting to note that equation
(\ref{mixedbc}) gives a quite natural hint of how ohmic losses on the
contour induce a small amount of complexity for wavefunctions that are
purely real in the unperturbed limit. As long as a perturbative approach is
valid, an immediate connection is deduced between the complex character of
wavefunctions and the fluctuating part of the widths embodied in the
quantity $\xi$ defined in (\ref{defxi})\cite{qfactor}.

\bibliography{smatrix}

\end{document}

%% file: antenne_pos2.pstex_t
\begin{picture}(0,0)%
\includegraphics{antenne_pos2.pstex}%
\end{picture}%
\setlength{\unitlength}{4144sp}%
\begingroup\makeatletter\ifx\SetFigFont\undefined%
\gdef\SetFigFont#1#2#3#4#5{%
  \reset@font\fontsize{#1}{#2pt}%
  \fontfamily{#3}\fontseries{#4}\fontshape{#5}%
  \selectfont}%
\fi\endgroup%
\begin{picture}(3122,1960)(317,-1455)
\put(722,-183){\makebox(0,0)[lb]{\smash{{\SetFigFont{10}{12.0}{\rmdefault}{\bfdefault}{\updefault}{\color[rgb]{0,0,0}10}%
}}}}
\put(2977,-1216){\makebox(0,0)[lb]{\smash{{\SetFigFont{10}{12.0}{\rmdefault}{\bfdefault}{\updefault}{\color[rgb]{0,0,0}1}%
}}}}
\put(2451,-525){\makebox(0,0)[lb]{\smash{{\SetFigFont{10}{12.0}{\rmdefault}{\bfdefault}{\updefault}{\color[rgb]{0,0,0}3}%
}}}}
\put(2639,-22){\makebox(0,0)[lb]{\smash{{\SetFigFont{10}{12.0}{\rmdefault}{\bfdefault}{\updefault}{\color[rgb]{0,0,0}4}%
}}}}
\put(1754,-448){\makebox(0,0)[lb]{\smash{{\SetFigFont{10}{12.0}{\rmdefault}{\bfdefault}{\updefault}{\color[rgb]{0,0,0}6}%
}}}}
\put(897,-873){\makebox(0,0)[lb]{\smash{{\SetFigFont{10}{12.0}{\rmdefault}{\bfdefault}{\updefault}{\color[rgb]{0,0,0}7}%
}}}}
\put(1141,-332){\makebox(0,0)[lb]{\smash{{\SetFigFont{10}{12.0}{\rmdefault}{\bfdefault}{\updefault}{\color[rgb]{0,0,0}8}%
}}}}
\put(1406, 56){\makebox(0,0)[lb]{\smash{{\SetFigFont{10}{12.0}{\rmdefault}{\bfdefault}{\updefault}{\color[rgb]{0,0,0}9}%
}}}}
\put(2264,-1024){\makebox(0,0)[lb]{\smash{{\SetFigFont{10}{12.0}{\rmdefault}{\bfdefault}{\updefault}{\color[rgb]{0,0,0}2}%
}}}}
\end{picture}%

%% file: antenne_model.pstex_t
\begin{picture}(0,0)%
\includegraphics{antenne_model.pstex}%
\end{picture}%
\setlength{\unitlength}{4144sp}%
\begingroup\makeatletter\ifx\SetFigFont\undefined%
\gdef\SetFigFont#1#2#3#4#5{%
  \reset@font\fontsize{#1}{#2pt}%
  \fontfamily{#3}\fontseries{#4}\fontshape{#5}%
  \selectfont}%
\fi\endgroup%
\begin{picture}(3164,2141)(744,-1735)
\put(3329,-1029){\makebox(0,0)[lb]{\smash{{\SetFigFont{10}{12.0}{\familydefault}{\mddefault}{\updefault}$0$}}}}
\put(3324,-329){\makebox(0,0)[lb]{\smash{{\SetFigFont{10}{12.0}{\familydefault}{\mddefault}{\updefault}$-l^*+l$}}}}
\put(3302,183){\makebox(0,0)[lb]{\smash{{\SetFigFont{10}{12.0}{\familydefault}{\mddefault}{\updefault}$-l^*$}}}}
\put(1000, 11){\makebox(0,0)[lb]{\smash{{\SetFigFont{10}{12.0}{\familydefault}{\mddefault}{\updefault}Cavity}}}}
\put(2088,-1695){\makebox(0,0)[lb]{\smash{{\SetFigFont{10}{12.0}{\familydefault}{\mddefault}{\updefault}$z$}}}}
\put(2368,-1272){\makebox(0,0)[b]{\smash{{\SetFigFont{10}{12.0}{\familydefault}{\mddefault}{\updefault}{\color[rgb]{0,0,0}TEM}%
}}}}
\put(1599,-660){\rotatebox{90.0}{\makebox(0,0)[b]{\smash{{\SetFigFont{10}{12.0}{\familydefault}{\mddefault}{\updefault}Perturbed}}}}}
\put(1761,-661){\rotatebox{90.0}{\makebox(0,0)[b]{\smash{{\SetFigFont{10}{12.0}{\familydefault}{\mddefault}{\updefault}Region}}}}}
\end{picture}%

%% file: test_transmission.pstex_t
\begin{picture}(0,0)%
\includegraphics{test_transmission.pstex}%
\end{picture}%
\setlength{\unitlength}{4144sp}%
\begingroup\makeatletter\ifx\SetFigFont\undefined%
\gdef\SetFigFont#1#2#3#4#5{%
  \reset@font\fontsize{#1}{#2pt}%
  \fontfamily{#3}\fontseries{#4}\fontshape{#5}%
  \selectfont}%
\fi\endgroup%
\begin{picture}(3400,2375)(1571,-2551)
\put(2168,-529){\makebox(0,0)[lb]{\smash{{\SetFigFont{10}{12.0}{\familydefault}{\mddefault}{\updefault}{\color[rgb]{0,0,0}$T^2\times 10^{-13}$\,m$^2$.s$^{-2}$}%
}}}}
\put(3183,-2511){\makebox(0,0)[b]{\smash{{\SetFigFont{10}{12.0}{\familydefault}{\mddefault}{\updefault}{\color[rgb]{0,0,0}Frequency in GHz}%
}}}}
\put(1725,-2249){\makebox(0,0)[rb]{\smash{{\SetFigFont{10}{12.0}{\familydefault}{\mddefault}{\updefault}0}}}}
\put(1725,-1447){\makebox(0,0)[rb]{\smash{{\SetFigFont{10}{12.0}{\familydefault}{\mddefault}{\updefault}8}}}}
\put(1725,-1047){\makebox(0,0)[rb]{\smash{{\SetFigFont{10}{12.0}{\familydefault}{\mddefault}{\updefault}12}}}}
\put(1725,-645){\makebox(0,0)[rb]{\smash{{\SetFigFont{10}{12.0}{\familydefault}{\mddefault}{\updefault}16}}}}
\put(1725,-1849){\makebox(0,0)[rb]{\smash{{\SetFigFont{10}{12.0}{\familydefault}{\mddefault}{\updefault}4}}}}
\put(2088,-2354){\makebox(0,0)[b]{\smash{{\SetFigFont{10}{12.0}{\familydefault}{\mddefault}{\updefault}1}}}}
\put(2722,-2354){\makebox(0,0)[b]{\smash{{\SetFigFont{10}{12.0}{\familydefault}{\mddefault}{\updefault}2}}}}
\put(3354,-2354){\makebox(0,0)[b]{\smash{{\SetFigFont{10}{12.0}{\familydefault}{\mddefault}{\updefault}3}}}}
\put(3987,-2354){\makebox(0,0)[b]{\smash{{\SetFigFont{10}{12.0}{\familydefault}{\mddefault}{\updefault}4}}}}
\put(4621,-2354){\makebox(0,0)[b]{\smash{{\SetFigFont{10}{12.0}{\familydefault}{\mddefault}{\updefault}5}}}}
\end{picture}%

%% file: fit_trans_pre.pstex_t
\begin{picture}(0,0)%
\includegraphics{fit_trans_pre.pstex}%
\end{picture}%
\setlength{\unitlength}{4144sp}%
\begingroup\makeatletter\ifx\SetFigFont\undefined%
\gdef\SetFigFont#1#2#3#4#5{%
  \reset@font\fontsize{#1}{#2pt}%
  \fontfamily{#3}\fontseries{#4}\fontshape{#5}%
  \selectfont}%
\fi\endgroup%
\begin{picture}(3170,4302)(2696,-5480)
\put(4179,-1573){\makebox(0,0)[lb]{\smash{{\SetFigFont{10}{12.0}{\familydefault}{\mddefault}{\updefault}{\color[rgb]{0,0,0}$2T^2(\omega_n)\vert\psi(\vec{r}_a)\vert^2$}%
}}}}
\put(3850,-5480){\makebox(0,0)[lb]{\smash{{\SetFigFont{10}{12.0}{\familydefault}{\mddefault}{\updefault}{\color[rgb]{0,0,0}antenna's index}%
}}}}
\put(2886,-3516){\makebox(0,0)[lb]{\smash{{\SetFigFont{10}{12.0}{\familydefault}{\mddefault}{\updefault}{\color[rgb]{0,0,0}$n=238$}%
}}}}
\put(3086,-5249){\makebox(0,0)[b]{\smash{{\SetFigFont{10}{12.0}{\familydefault}{\mddefault}{\updefault} 2}}}}
\put(3422,-5249){\makebox(0,0)[b]{\smash{{\SetFigFont{10}{12.0}{\familydefault}{\mddefault}{\updefault} 3}}}}
\put(2750,-5249){\makebox(0,0)[b]{\smash{{\SetFigFont{10}{12.0}{\familydefault}{\mddefault}{\updefault} 1}}}}
\put(3758,-5249){\makebox(0,0)[b]{\smash{{\SetFigFont{10}{12.0}{\familydefault}{\mddefault}{\updefault} 4}}}}
\put(4095,-5249){\makebox(0,0)[b]{\smash{{\SetFigFont{10}{12.0}{\familydefault}{\mddefault}{\updefault} 5}}}}
\put(4430,-5249){\makebox(0,0)[b]{\smash{{\SetFigFont{10}{12.0}{\familydefault}{\mddefault}{\updefault} 6}}}}
\put(4766,-5249){\makebox(0,0)[b]{\smash{{\SetFigFont{10}{12.0}{\familydefault}{\mddefault}{\updefault} 7}}}}
\put(5102,-5249){\makebox(0,0)[b]{\smash{{\SetFigFont{10}{12.0}{\familydefault}{\mddefault}{\updefault} 8}}}}
\put(5439,-5249){\makebox(0,0)[b]{\smash{{\SetFigFont{10}{12.0}{\familydefault}{\mddefault}{\updefault} 9}}}}
\put(5774,-5249){\makebox(0,0)[b]{\smash{{\SetFigFont{10}{12.0}{\familydefault}{\mddefault}{\updefault} 10}}}}
\put(2886,-1603){\makebox(0,0)[lb]{\smash{{\SetFigFont{10}{12.0}{\familydefault}{\mddefault}{\updefault}{\color[rgb]{0,0,0}$n=78$}%
}}}}
\end{picture}%

%% file: fit_trans2_lin_fin.pstex_t
\begin{picture}(0,0)%
\includegraphics{fit_trans2_lin_fin.pstex}%
\end{picture}%
\setlength{\unitlength}{4144sp}%
\begingroup\makeatletter\ifx\SetFigFont\undefined%
\gdef\SetFigFont#1#2#3#4#5{%
  \reset@font\fontsize{#1}{#2pt}%
  \fontfamily{#3}\fontseries{#4}\fontshape{#5}%
  \selectfont}%
\fi\endgroup%
\begin{picture}(3716,2714)(1385,-2688)
\put(2554,-1730){\makebox(0,0)[b]{\smash{{\SetFigFont{10}{12.0}{\familydefault}{\mddefault}{\updefault}{\color[rgb]{0,0,0}2.4}%
}}}}
\put(3382,-1730){\makebox(0,0)[b]{\smash{{\SetFigFont{10}{12.0}{\familydefault}{\mddefault}{\updefault}{\color[rgb]{0,0,0}2.8}%
}}}}
\put(1726,-1730){\makebox(0,0)[b]{\smash{{\SetFigFont{10}{12.0}{\familydefault}{\mddefault}{\updefault}{\color[rgb]{0,0,0}2}%
}}}}
\put(1846,-376){\makebox(0,0)[lb]{\smash{{\SetFigFont{10}{12.0}{\familydefault}{\mddefault}{\updefault}{\color[rgb]{0,0,0}$2T^2(\omega_n)\langle\vert\psi(\vec{r}_a)\vert^2\rangle_a$}%
}}}}
\put(3882,-2389){\makebox(0,0)[b]{\smash{{\SetFigFont{10}{12.0}{\familydefault}{\mddefault}{\updefault}{\color[rgb]{0,0,0}4}%
}}}}
\put(3291,-2389){\makebox(0,0)[b]{\smash{{\SetFigFont{10}{12.0}{\familydefault}{\mddefault}{\updefault}{\color[rgb]{0,0,0}3}%
}}}}
\put(2700,-2389){\makebox(0,0)[b]{\smash{{\SetFigFont{10}{12.0}{\familydefault}{\mddefault}{\updefault}{\color[rgb]{0,0,0}2}%
}}}}
\put(2109,-2389){\makebox(0,0)[b]{\smash{{\SetFigFont{10}{12.0}{\familydefault}{\mddefault}{\updefault}{\color[rgb]{0,0,0}1}%
}}}}
\put(1519,-2389){\makebox(0,0)[b]{\smash{{\SetFigFont{10}{12.0}{\familydefault}{\mddefault}{\updefault}{\color[rgb]{0,0,0}0}%
}}}}
\put(5063,-2389){\makebox(0,0)[b]{\smash{{\SetFigFont{10}{12.0}{\familydefault}{\mddefault}{\updefault}{\color[rgb]{0,0,0}6}%
}}}}
\put(1424,-79){\makebox(0,0)[b]{\smash{{\SetFigFont{10}{12.0}{\familydefault}{\mddefault}{\updefault}{\color[rgb]{0,0,0}7}%
}}}}
\put(1424,-392){\makebox(0,0)[b]{\smash{{\SetFigFont{10}{12.0}{\familydefault}{\mddefault}{\updefault}{\color[rgb]{0,0,0}6}%
}}}}
\put(1424,-702){\makebox(0,0)[b]{\smash{{\SetFigFont{10}{12.0}{\familydefault}{\mddefault}{\updefault}{\color[rgb]{0,0,0}5}%
}}}}
\put(1424,-1014){\makebox(0,0)[b]{\smash{{\SetFigFont{10}{12.0}{\familydefault}{\mddefault}{\updefault}{\color[rgb]{0,0,0}4}%
}}}}
\put(1424,-1325){\makebox(0,0)[b]{\smash{{\SetFigFont{10}{12.0}{\familydefault}{\mddefault}{\updefault}{\color[rgb]{0,0,0}3}%
}}}}
\put(1424,-1637){\makebox(0,0)[b]{\smash{{\SetFigFont{10}{12.0}{\familydefault}{\mddefault}{\updefault}{\color[rgb]{0,0,0}2}%
}}}}
\put(1424,-1948){\makebox(0,0)[b]{\smash{{\SetFigFont{10}{12.0}{\familydefault}{\mddefault}{\updefault}{\color[rgb]{0,0,0}1}%
}}}}
\put(1424,-2260){\makebox(0,0)[b]{\smash{{\SetFigFont{10}{12.0}{\familydefault}{\mddefault}{\updefault}{\color[rgb]{0,0,0}0}%
}}}}
\put(3293,-2648){\makebox(0,0)[b]{\smash{{\SetFigFont{10}{12.0}{\familydefault}{\mddefault}{\updefault}Frequency in GHz}}}}
\put(4472,-2389){\makebox(0,0)[b]{\smash{{\SetFigFont{10}{12.0}{\familydefault}{\mddefault}{\updefault}{\color[rgb]{0,0,0}5}%
}}}}
\end{picture}%

%% file: xi_impulse.pstex_t
\begin{picture}(0,0)%
\includegraphics{xi_impulse.pstex}%
\end{picture}%
\setlength{\unitlength}{4144sp}%
\begingroup\makeatletter\ifx\SetFigFont\undefined%
\gdef\SetFigFont#1#2#3#4#5{%
  \reset@font\fontsize{#1}{#2pt}%
  \fontfamily{#3}\fontseries{#4}\fontshape{#5}%
  \selectfont}%
\fi\endgroup%
\begin{picture}(3491,2227)(1218,-2494)
\put(1350,-1189){\rotatebox{90.0}{\makebox(0,0)[b]{\smash{{\SetFigFont{10}{12.0}{\familydefault}{\mddefault}{\updefault}{\color[rgb]{0,0,0}$\xi_{l,m}-1$}%
}}}}}
\put(3192,-2454){\makebox(0,0)[b]{\smash{{\SetFigFont{10}{12.0}{\familydefault}{\mddefault}{\updefault}Frequency in GHz}}}}
\put(4178,-2278){\makebox(0,0)[b]{\smash{{\SetFigFont{10}{12.0}{\familydefault}{\mddefault}{\updefault}5}}}}
\put(4671,-2278){\makebox(0,0)[b]{\smash{{\SetFigFont{10}{12.0}{\familydefault}{\mddefault}{\updefault}6}}}}
\put(3192,-2278){\makebox(0,0)[b]{\smash{{\SetFigFont{10}{12.0}{\familydefault}{\mddefault}{\updefault}3}}}}
\put(2205,-2278){\makebox(0,0)[b]{\smash{{\SetFigFont{10}{12.0}{\familydefault}{\mddefault}{\updefault}1}}}}
\put(1650,-1952){\makebox(0,0)[rb]{\smash{{\SetFigFont{10}{12.0}{\familydefault}{\mddefault}{\updefault}-0.2}}}}
\put(1650,-1588){\makebox(0,0)[rb]{\smash{{\SetFigFont{10}{12.0}{\familydefault}{\mddefault}{\updefault}-0.1}}}}
\put(1650,-1224){\makebox(0,0)[rb]{\smash{{\SetFigFont{10}{12.0}{\familydefault}{\mddefault}{\updefault}0}}}}
\put(1650,-860){\makebox(0,0)[rb]{\smash{{\SetFigFont{10}{12.0}{\familydefault}{\mddefault}{\updefault}0.1}}}}
\put(1650,-496){\makebox(0,0)[rb]{\smash{{\SetFigFont{10}{12.0}{\familydefault}{\mddefault}{\updefault}0.2}}}}
\put(2699,-2278){\makebox(0,0)[b]{\smash{{\SetFigFont{10}{12.0}{\familydefault}{\mddefault}{\updefault}2}}}}
\put(1713,-2278){\makebox(0,0)[b]{\smash{{\SetFigFont{10}{12.0}{\familydefault}{\mddefault}{\updefault}0}}}}
\put(3685,-2278){\makebox(0,0)[b]{\smash{{\SetFigFont{10}{12.0}{\familydefault}{\mddefault}{\updefault}4}}}}
\end{picture}%

%% file: largeurs_vide_ins.pstex_t
\begin{picture}(0,0)%
\includegraphics{largeurs_vide_ins.pstex}%
\end{picture}%
\setlength{\unitlength}{4144sp}%
\begingroup\makeatletter\ifx\SetFigFont\undefined%
\gdef\SetFigFont#1#2#3#4#5{%
  \reset@font\fontsize{#1}{#2pt}%
  \fontfamily{#3}\fontseries{#4}\fontshape{#5}%
  \selectfont}%
\fi\endgroup%
\begin{picture}(3312,2380)(2477,-3382)
\put(4191,-3342){\makebox(0,0)[b]{\smash{{\SetFigFont{10}{12.0}{\familydefault}{\mddefault}{\updefault}{\color[rgb]{0,0,0}Frequency in GHz}%
}}}}
\put(2881,-1456){\makebox(0,0)[lb]{\smash{{\SetFigFont{10}{12.0}{\familydefault}{\mddefault}{\updefault}{\color[rgb]{0,0,0}$\Gamma_n^{\textrm{ohm}}/2\pi$ in MHz}%
}}}}
\put(2601,-1352){\makebox(0,0)[rb]{\smash{{\SetFigFont{10}{12.0}{\familydefault}{\mddefault}{\updefault}.9}}}}
\put(4698,-3178){\makebox(0,0)[b]{\smash{{\SetFigFont{10}{12.0}{\familydefault}{\mddefault}{\updefault} 2}}}}
\put(5212,-3178){\makebox(0,0)[b]{\smash{{\SetFigFont{10}{12.0}{\familydefault}{\mddefault}{\updefault} 2.5}}}}
\put(3149,-3178){\makebox(0,0)[b]{\smash{{\SetFigFont{10}{12.0}{\familydefault}{\mddefault}{\updefault} 0.5}}}}
\put(3664,-3178){\makebox(0,0)[b]{\smash{{\SetFigFont{10}{12.0}{\familydefault}{\mddefault}{\updefault} 1}}}}
\put(5736,-3178){\makebox(0,0)[b]{\smash{{\SetFigFont{10}{12.0}{\familydefault}{\mddefault}{\updefault} 3}}}}
\put(2646,-3178){\makebox(0,0)[b]{\smash{{\SetFigFont{10}{12.0}{\familydefault}{\mddefault}{\updefault} 0}}}}
\put(2601,-1597){\makebox(0,0)[rb]{\smash{{\SetFigFont{10}{12.0}{\familydefault}{\mddefault}{\updefault}.8}}}}
\put(2601,-2820){\makebox(0,0)[rb]{\smash{{\SetFigFont{10}{12.0}{\familydefault}{\mddefault}{\updefault}.3}}}}
\put(2601,-2331){\makebox(0,0)[rb]{\smash{{\SetFigFont{10}{12.0}{\familydefault}{\mddefault}{\updefault}.5}}}}
\put(2601,-3065){\makebox(0,0)[rb]{\smash{{\SetFigFont{10}{12.0}{\familydefault}{\mddefault}{\updefault}.2}}}}
\put(2601,-2086){\makebox(0,0)[rb]{\smash{{\SetFigFont{10}{12.0}{\familydefault}{\mddefault}{\updefault}.6}}}}
\put(2601,-1841){\makebox(0,0)[rb]{\smash{{\SetFigFont{10}{12.0}{\familydefault}{\mddefault}{\updefault}.7}}}}
\put(2601,-1107){\makebox(0,0)[rb]{\smash{{\SetFigFont{10}{12.0}{\familydefault}{\mddefault}{\updefault} 1}}}}
\put(2601,-2575){\makebox(0,0)[rb]{\smash{{\SetFigFont{10}{12.0}{\familydefault}{\mddefault}{\updefault}.4}}}}
\put(4183,-3178){\makebox(0,0)[b]{\smash{{\SetFigFont{10}{12.0}{\familydefault}{\mddefault}{\updefault} 1.5}}}}
\end{picture}%

%% file: smatrix.bbl
\begin{thebibliography}{25}
\expandafter\ifx\csname natexlab\endcsname\relax\def\natexlab#1{#1}\fi
\expandafter\ifx\csname bibnamefont\endcsname\relax
  \def\bibnamefont#1{#1}\fi
\expandafter\ifx\csname bibfnamefont\endcsname\relax
  \def\bibfnamefont#1{#1}\fi
\expandafter\ifx\csname citenamefont\endcsname\relax
  \def\citenamefont#1{#1}\fi
\expandafter\ifx\csname url\endcsname\relax
  \def\url#1{\texttt{#1}}\fi
\expandafter\ifx\csname urlprefix\endcsname\relax\def\urlprefix{URL }\fi
\providecommand{\bibinfo}[2]{#2}
\providecommand{\eprint}[2][]{\url{#2}}

\bibitem[{\citenamefont{St{\"o}ckmann}(1999)}]{Stoeckmannbook}
\bibinfo{author}{\bibfnamefont{H.-J.} \bibnamefont{St{\"o}ckmann}},
  \emph{\bibinfo{title}{Quantum Chaos}} (\bibinfo{publisher}{Cambridge
  University Press}, \bibinfo{year}{1999}).

\bibitem[{\citenamefont{St{\"o}ckmann and Stein}(1990)}]{Stoeckmann_1990}
\bibinfo{author}{\bibfnamefont{H.-J.} \bibnamefont{St{\"o}ckmann}}
  \bibnamefont{and} \bibinfo{author}{\bibfnamefont{J.}~\bibnamefont{Stein}},
  \bibinfo{journal}{Phys. Rev. Lett.} \textbf{\bibinfo{volume}{64}},
  \bibinfo{pages}{2215} (\bibinfo{year}{1990}).

\bibitem[{\citenamefont{Gr{\"a}f et~al.}(1992)\citenamefont{Gr{\"a}f, Harney,
  Lengeler, Lewenkopf, Rangacharyulu, Richter, Schardt, and
  Weidenm{\"u}ller}}]{Graef_1992}
\bibinfo{author}{\bibfnamefont{H.-D.} \bibnamefont{Gr{\"a}f}},
  \bibinfo{author}{\bibfnamefont{H.}~\bibnamefont{Harney}},
  \bibinfo{author}{\bibfnamefont{H.}~\bibnamefont{Lengeler}},
  \bibinfo{author}{\bibfnamefont{C.}~\bibnamefont{Lewenkopf}},
  \bibinfo{author}{\bibfnamefont{C.}~\bibnamefont{Rangacharyulu}},
  \bibinfo{author}{\bibfnamefont{A.}~\bibnamefont{Richter}},
  \bibinfo{author}{\bibfnamefont{P.}~\bibnamefont{Schardt}}, \bibnamefont{and}
  \bibinfo{author}{\bibfnamefont{H.}~\bibnamefont{Weidenm{\"u}ller}},
  \bibinfo{journal}{Phys. Rev. Lett.} \textbf{\bibinfo{volume}{69}},
  \bibinfo{pages}{1296} (\bibinfo{year}{1992}).

\bibitem[{\citenamefont{Haake et~al.}(1991)\citenamefont{Haake, Lenz, \v{S}eba,
  Stein, St{\"o}ckmann, and \.{Z}yczkowski}}]{Haake_1991}
\bibinfo{author}{\bibfnamefont{F.}~\bibnamefont{Haake}},
  \bibinfo{author}{\bibfnamefont{G.}~\bibnamefont{Lenz}},
  \bibinfo{author}{\bibfnamefont{P.}~\bibnamefont{\v{S}eba}},
  \bibinfo{author}{\bibfnamefont{J.}~\bibnamefont{Stein}},
  \bibinfo{author}{\bibfnamefont{H.-J.} \bibnamefont{St{\"o}ckmann}},
  \bibnamefont{and}
  \bibinfo{author}{\bibfnamefont{K.}~\bibnamefont{\.{Z}yczkowski}},
  \bibinfo{journal}{Phys. Rev. A} \textbf{\bibinfo{volume}{44}},
  \bibinfo{pages}{R6161} (\bibinfo{year}{1991}).

\bibitem[{\citenamefont{Alt et~al.}(1995)\citenamefont{Alt, Gr{\"a}f, Harney,
  Lengeler, Richter, Schardt, and Weidenm{\"u}ller}}]{alt_1995}
\bibinfo{author}{\bibfnamefont{H.}~\bibnamefont{Alt}},
  \bibinfo{author}{\bibfnamefont{H.-D.} \bibnamefont{Gr{\"a}f}},
  \bibinfo{author}{\bibfnamefont{H.}~\bibnamefont{Harney}},
  \bibinfo{author}{\bibfnamefont{H.}~\bibnamefont{Lengeler}},
  \bibinfo{author}{\bibfnamefont{A.}~\bibnamefont{Richter}},
  \bibinfo{author}{\bibfnamefont{P.}~\bibnamefont{Schardt}}, \bibnamefont{and}
  \bibinfo{author}{\bibfnamefont{H.}~\bibnamefont{Weidenm{\"u}ller}},
  \bibinfo{journal}{Phys. Rev. Lett.} \textbf{\bibinfo{volume}{74}},
  \bibinfo{pages}{62} (\bibinfo{year}{1995}).

\bibitem[{\citenamefont{Ericson}(1963)}]{Ericson_1963}
\bibinfo{author}{\bibfnamefont{T.}~\bibnamefont{Ericson}},
  \bibinfo{journal}{Ann. Phys.} \textbf{\bibinfo{volume}{23}},
  \bibinfo{pages}{390} (\bibinfo{year}{1963}).

\bibitem[{\citenamefont{Schr{\oe}der}(1962)}]{Schroeder_1962b}
\bibinfo{author}{\bibfnamefont{M.}~\bibnamefont{Schr{\oe}der}},
  \bibinfo{journal}{J. Acoust. Soc. Am.} \textbf{\bibinfo{volume}{34}},
  \bibinfo{pages}{1819} (\bibinfo{year}{1962}).

\bibitem[{\citenamefont{Verbaarschot et~al.}(1985)\citenamefont{Verbaarschot,
  Weidenm{\"u}ller, and Zirnbauer}}]{Verbaarschot_1985}
\bibinfo{author}{\bibfnamefont{J.}~\bibnamefont{Verbaarschot}},
  \bibinfo{author}{\bibfnamefont{H.}~\bibnamefont{Weidenm{\"u}ller}},
  \bibnamefont{and}
  \bibinfo{author}{\bibfnamefont{M.}~\bibnamefont{Zirnbauer}},
  \bibinfo{journal}{Phys. Rep.} \textbf{\bibinfo{volume}{129}},
  \bibinfo{pages}{367} (\bibinfo{year}{1985}).

\bibitem[{\citenamefont{Sch{\"a}fer et~al.}(2003)\citenamefont{Sch{\"a}fer,
  Gorin, Seligman, and St{\"o}ckmann}}]{Schafer_2003}
\bibinfo{author}{\bibfnamefont{R.}~\bibnamefont{Sch{\"a}fer}},
  \bibinfo{author}{\bibfnamefont{T.}~\bibnamefont{Gorin}},
  \bibinfo{author}{\bibfnamefont{T.}~\bibnamefont{Seligman}}, \bibnamefont{and}
  \bibinfo{author}{\bibfnamefont{H.-J.} \bibnamefont{St{\"o}ckmann}},
  \bibinfo{journal}{J. Phys. A: Math. Gen.} \textbf{\bibinfo{volume}{36}},
  \bibinfo{pages}{3289} (\bibinfo{year}{2003}).

\bibitem[{\citenamefont{Rozhkov et~al.}(2003)\citenamefont{Rozhkov, Fyodorov,
  and Weaver}}]{Rozhkov_2003}
\bibinfo{author}{\bibfnamefont{I.}~\bibnamefont{Rozhkov}},
  \bibinfo{author}{\bibfnamefont{Y.~V.} \bibnamefont{Fyodorov}},
  \bibnamefont{and} \bibinfo{author}{\bibfnamefont{R.~L.}
  \bibnamefont{Weaver}}, \bibinfo{journal}{Phys. Rev. E}
  \textbf{\bibinfo{volume}{68}}, \bibinfo{pages}{016204}
  (\bibinfo{year}{2003}).

\bibitem[{\citenamefont{Dittes}(2000)}]{Dittes_2000}
\bibinfo{author}{\bibfnamefont{F.-M.} \bibnamefont{Dittes}},
  \bibinfo{journal}{Phys. Rep.} \textbf{\bibinfo{volume}{339}},
  \bibinfo{pages}{215} (\bibinfo{year}{2000}).

\bibitem[{\citenamefont{Harrington}(2001)}]{Harringtonbook}
\bibinfo{author}{\bibfnamefont{R.~F.} \bibnamefont{Harrington}},
  \emph{\bibinfo{title}{Time-Harmonic Electromagnetic Fields}}, A Classic
  Reissue (\bibinfo{publisher}{Wiley-Interscience}, \bibinfo{year}{2001}).

\bibitem[{\citenamefont{Collin}(1991)}]{Collinbook}
\bibinfo{author}{\bibfnamefont{R.~E.} \bibnamefont{Collin}},
  \emph{\bibinfo{title}{Field Theory of Guided Waves}}, Electromagnetic waves
  (\bibinfo{publisher}{IEEE Press}, \bibinfo{address}{New York},
  \bibinfo{year}{1991}), \bibinfo{edition}{2nd} ed.

\bibitem[{\citenamefont{Liu and Grimes}(2000)}]{Liu_2000}
\bibinfo{author}{\bibfnamefont{G.}~\bibnamefont{Liu}} \bibnamefont{and}
  \bibinfo{author}{\bibfnamefont{C.~A.} \bibnamefont{Grimes}},
  \bibinfo{journal}{Microwave Opt. Technol. Lett.}
  \textbf{\bibinfo{volume}{26}}, \bibinfo{pages}{30} (\bibinfo{year}{2000}).

\bibitem[{\citenamefont{Shen and MacPhie}(1996)}]{Shen_1996}
\bibinfo{author}{\bibfnamefont{Z.}~\bibnamefont{Shen}} \bibnamefont{and}
  \bibinfo{author}{\bibfnamefont{R.~H.} \bibnamefont{MacPhie}},
  \bibinfo{journal}{Radio Sci.} \textbf{\bibinfo{volume}{31}},
  \bibinfo{pages}{1037} (\bibinfo{year}{1996}).

\bibitem[{\citenamefont{Eom et~al.}(2000)\citenamefont{Eom, Cho, and
  Kwon}}]{Eom_2000}
\bibinfo{author}{\bibfnamefont{H.~J.} \bibnamefont{Eom}},
  \bibinfo{author}{\bibfnamefont{Y.~H.} \bibnamefont{Cho}}, \bibnamefont{and}
  \bibinfo{author}{\bibfnamefont{M.~S.} \bibnamefont{Kwon}},
  \bibinfo{journal}{IEEE Trans. Antennas Propag.}
  \textbf{\bibinfo{volume}{48}}, \bibinfo{pages}{1142} (\bibinfo{year}{2000}).

\bibitem[{\citenamefont{Feshbach}(1992)}]{Feshbachbook}
\bibinfo{author}{\bibfnamefont{H.}~\bibnamefont{Feshbach}},
  \emph{\bibinfo{title}{Theoretical Nuclear Physics}} (\bibinfo{publisher}{John
  Wiley {\&} Sons}, \bibinfo{year}{1992}).

\bibitem[{\citenamefont{Fyodorov and Sommers}(1997)}]{Fyodorov_1997}
\bibinfo{author}{\bibfnamefont{Y.}~\bibnamefont{Fyodorov}} \bibnamefont{and}
  \bibinfo{author}{\bibfnamefont{H.-J.} \bibnamefont{Sommers}},
  \bibinfo{journal}{J. Math. Phys.} \textbf{\bibinfo{volume}{38}},
  \bibinfo{pages}{1918} (\bibinfo{year}{1997}).

\bibitem[{\citenamefont{Albeverio et~al.}(1996)\citenamefont{Albeverio, Haake,
  Kurasov, Ku{\'s}, and \v{S}eba}}]{Albeverio_1996}
\bibinfo{author}{\bibfnamefont{S.}~\bibnamefont{Albeverio}},
  \bibinfo{author}{\bibfnamefont{F.}~\bibnamefont{Haake}},
  \bibinfo{author}{\bibfnamefont{P.}~\bibnamefont{Kurasov}},
  \bibinfo{author}{\bibfnamefont{M.}~\bibnamefont{Ku{\'s}}}, \bibnamefont{and}
  \bibinfo{author}{\bibfnamefont{P.}~\bibnamefont{\v{S}eba}},
  \bibinfo{journal}{J. Math. Phys.} \textbf{\bibinfo{volume}{37}},
  \bibinfo{pages}{4888} (\bibinfo{year}{1996}).

\bibitem[{\citenamefont{Jackson}(1965)}]{Jacksonbook}
\bibinfo{author}{\bibfnamefont{J.~D.} \bibnamefont{Jackson}},
  \emph{\bibinfo{title}{Classical Electrodynamics}}
  (\bibinfo{publisher}{Academic Press}, \bibinfo{address}{New York},
  \bibinfo{year}{1965}).

\bibitem[{\citenamefont{Barth{\'e}lemy
  et~al.}(2004)\citenamefont{Barth{\'e}lemy, Legrand, and
  Mortessagne}}]{qfactor}
\bibinfo{author}{\bibfnamefont{J.}~\bibnamefont{Barth{\'e}lemy}},
  \bibinfo{author}{\bibfnamefont{O.}~\bibnamefont{Legrand}}, \bibnamefont{and}
  \bibinfo{author}{\bibfnamefont{F.}~\bibnamefont{Mortessagne}}
  (\bibinfo{year}{2004}), \eprint{cond-mat/0402029}.

\bibitem[{\citenamefont{Barth{\'e}lemy}(2003)}]{Jerome_2003}
\bibinfo{author}{\bibfnamefont{J.}~\bibnamefont{Barth{\'e}lemy}}, Ph.D. thesis,
  \bibinfo{school}{Universit{\'e} Paris 7 - Denis Diderot}
  (\bibinfo{year}{2003}),
  \urlprefix\url{http://tel.ccsd.cnrs.fr/documents/archives0/00/00/41/14/}.

\bibitem[{\citenamefont{Kuhl et~al.}(2000)\citenamefont{Kuhl, Persson, Barth,
  and St{\"o}ckmann}}]{Kuhl_2000}
\bibinfo{author}{\bibfnamefont{U.}~\bibnamefont{Kuhl}},
  \bibinfo{author}{\bibfnamefont{E.}~\bibnamefont{Persson}},
  \bibinfo{author}{\bibfnamefont{M.}~\bibnamefont{Barth}}, \bibnamefont{and}
  \bibinfo{author}{\bibfnamefont{H.-J.} \bibnamefont{St{\"o}ckmann}},
  \bibinfo{journal}{Eur. Phys. J. B} \textbf{\bibinfo{volume}{17}},
  \bibinfo{pages}{253} (\bibinfo{year}{2000}).

\bibitem[{\citenamefont{M{\'e}ndez-S{\'a}nchez
  et~al.}(2003)\citenamefont{M{\'e}ndez-S{\'a}nchez, U., Barth, Lewenkopf, and
  St{\"o}ckmann}}]{Mendez_2003}
\bibinfo{author}{\bibfnamefont{R.~A.} \bibnamefont{M{\'e}ndez-S{\'a}nchez}},
  \bibinfo{author}{\bibfnamefont{K.}~\bibnamefont{U.}},
  \bibinfo{author}{\bibfnamefont{M.}~\bibnamefont{Barth}},
  \bibinfo{author}{\bibfnamefont{H.}~\bibnamefont{Lewenkopf}},
  \bibnamefont{and} \bibinfo{author}{\bibfnamefont{H.-J.}
  \bibnamefont{St{\"o}ckmann}}, \bibinfo{journal}{Phys. Rev. Lett.}
  \textbf{\bibinfo{volume}{91}}, \bibinfo{pages}{174102}
  (\bibinfo{year}{2003}).

\bibitem[{\citenamefont{Savin and Sommers}(2003)}]{Savin_2003}
\bibinfo{author}{\bibfnamefont{D.~V.} \bibnamefont{Savin}} \bibnamefont{and}
  \bibinfo{author}{\bibfnamefont{H.-J.} \bibnamefont{Sommers}},
  \bibinfo{journal}{Phys. Rev. E} \textbf{\bibinfo{volume}{68}},
  \bibinfo{pages}{036211} (\bibinfo{year}{2003}).

\end{thebibliography}
